\documentclass{aa}  

\usepackage{afterpage}
\usepackage{graphicx}
\usepackage{multirow}
\usepackage{booktabs}
\usepackage{orcidlink}

\usepackage{txfonts}
\usepackage{mathtools}

\usepackage{pifont}

\usepackage{hyperref}
\hypersetup{
    colorlinks=true,       
    citecolor=blue,         
    linkcolor=blue,        
    urlcolor=blue       
}
\begin{document}

   \title{Close-in faint companions mimicking interferometric hot exozodiacal dust observations}
    \titlerunning{Faint companions mimicking hot exozodiacal dust}

\author{
    Katsiaryna Tsishchankava\orcidlink{0009-0002-9371-0740}\inst{1} 
    \and
    Florian Kirchschlager\orcidlink{0000-0002-3036-0184}\inst{2}
    \and
    Anton Krieger\orcidlink{0000-0002-3639-2435}\inst{1} 
    \and
    Thomas A. Stuber\orcidlink{0000-0003-2185-0525}\inst{3}
    \and
    Sebastian Wolf\orcidlink{0000-0001-7841-3452}\inst{1}
}

\institute{
    Institut für Theoretische Physik und Astrophysik, Christian-Albrechts-Universität zu Kiel, Leibnizstraße 15, 24118 Kiel, Germany\\
    \email{kattsish@astrophysik.uni-kiel.de} 
    \and
    Sterrenkundig Observatorium, Ghent University, Krijgslaan 281-S9, B-9000 Gent, Belgium
    \and
    Department of Astronomy and Steward Observatory, The University of Arizona, 933 North Cherry Ave, Tucson, AZ 85721, USA
}

   \date{Received April 2, 2025 / Accepted October 6,2025}

  \abstract
   { \textit{Context:} Interferometric observations of various nearby main sequence stars display an unexpected infrared excess, raising questions about its origin. The two dominant interpretations favor hot exozodiacal dust or a faint companion, both with the capacity to produce similar infrared interferometric signatures. 

    \textit{Method:} We modeled a system consisting of a limb-darkened star and a faint companion within a field of view of \( 2 \, \mathrm{au} \times 2 \, \mathrm{au} \), corresponding to an angular separation of up to \( 0.07 \)\,as from the star. We calculated the visibility and closure phases for three VLTI instruments (PIONIER, GRAVITY, and MATISSE), along with four telescope configurations (small, medium, large, and extended).

    \textit{Aim:} We aim to investigate the interferometric signatures of faint companions in the presence of a limb-darkened star and assess their detectability based on visibility and closure phase measurements. By modeling the VLTI instruments and telescope configurations, we explore the limitations of current detection methods and evaluate the challenges in distinguishing between hot exozodiacal dust and a faint companion as the source of the observed infrared excess.
    
    \textit{Results:} We derived an upper limit for the companion-induced visibility deficit of $\left|\Delta V(f)\right| \leq 2f$ for a companion-to-star flux ratios of $f \leq 10\,\%$, as well as upper limits on the closure phase, which changes linearly with $f$ and is inversely proportional to the visibility of the star. Contrary to the common interpretation that near-zero closure phases rule out the presence of a companion, we show that companions can remain undetected in closure phase data, as indicated by significant non-detection probabilities. However, these companions can still produce measurable visibility deficits that can even approach the theoretical upper limit. We confirmed our results by reevaluating an L-band observation of $\kappa$~Tuc~A using MATISSE. We found indications of a faint companion with a flux ratio of $0.7\,\%$ and an estimated non-detection probability of around $\sim21\,\%$, which could explain the variability of the previously observed visibility deficit. 

   \textit{Conclusions:}
    Previously applied companion rejection criteria, such as near-zero closure phases and flux estimates based on Gaussian-distributed dust densities, are not universally valid. This highlights the need for a reevaluation of companion rejections in former studies of the hot exozodiacal dust phenomenon. In addition, we propose a novel method for distinguishing both sources of the visibility deficit.
}

   \maketitle

\section{Introduction}

The detection and characterization of planetary systems around nearby main sequence stars is a central goal of modern astrophysics. However, studying the inner regions of these systems presents significant challenges due to their small angular separations from the host star and the high brightness contrast between the star and potential sources in its immediate environment. Infrared (IR) interferometry provides an effective method to address these challenges, offering angular resolutions of a few milli-arcseconds (mas).  Within a distance of $\sim 100\,$pc, this allows for the detection and resolution of structures within up to a tenth of an astronomical unit (au) from the central star. While interferometric techniques have made significant progress in identifying circumstellar dust structures and faint companions around main-sequence stars \citep[e.g.,][]{2011A&A...535A..68A,2014A&A...570A.127M,2019A&A...623L..11G,2020A&A...642L...2N}, distinguishing between these two phenomena remains a challenge. In particular, the near- and mid-infrared (NIR and MIR) signatures of close-in stellar or sub-stellar companions can mimic the presence of hot exozodiacal dust  (i.e., small dust grains residing within 1\,au from their host star). Likewise, the faint thermal and scattered-light signals produced by a small population of hot dust grains can seriously affect our ability to detect companions, including those orbiting within the habitable zone of their host stars \citep[e.g.,][]{2012PASP..124..799R,2021A&A...651A..45A,2023A&A...678A.121S,2023A&A...677A.187O}.

Since the first detection of a visibility deficit (i.e., a drop in interferometric fringe contrast caused by circumstellar radiation that is not fully accounted for by a stellar photosphere) around Vega \citep{2006via..conf..251A}, subsequent discoveries have revealed similar phenomena around main sequence stars of various spectral types \citep[e.g.,][]{2008A&A...487.1041A,Ertel2014,2009ApJ...704..150A,2013A&A...555A.104A,2017A&A...608A.113N,2021A&A...651A..45A,2025A&A...699A.144O}. These visibility deficits display values in the range of 0.01--0.05, observed with instruments such as CHARA/FLUOR in the K band \citep{2003SPIE.4838..280C} and the Very Large Telescope Interferometer (VLTI) in the H, K, and L bands. Furthermore, these deficits cannot be readily explained by the known outer debris disks. The leading explanation for this phenomenon is the presence of hot exozodiacal dust located in a narrow hot dust ring within <1\,au near the sublimation radius of a star \citep[e.g.,][]{2016ApJ...816...50R,2017MNRAS.467.1614K}. Further analysis suggests that while the dust primarily consists of small grains, a significant fraction of larger dust particles might also be present \citep{2023A&A...678A.121S}.
However, the mechanism responsible for this phenomenon remains unknown, as small, hot grains are expected to undergo rapid sublimation or be expelled by stellar radiation pressure. Local replenishment through steady-state collisional cascades is considered highly unlikely \citep{2007ApJ...658..569W, 2013A&A...555A.146L}. To account for the persistence of hot dust, several alternative mechanisms have been proposed, including a continuous resupply from distant regions of the system and dust-trapping processes near the star \citep{KRIVOV1998311,2008Icar..195..871K,2009Icar..201..395K,2016ApJ...816...50R,2019A&A...626A...2S,2020MNRAS.498.2798P,2020P&SS..18304581K,2025PASP..137c1001E}. However, none of the existing models have successfully reproduced the observed properties, coupled with the widespread occurrence of hot exozodiacal dust across different stellar types and evolutionary stages \citep{2022MNRAS.517.1436P}.

Further observations have highlighted the variability in the detected flux, including periods with no apparent infrared excess, that is, where no deficit was measured in the visibility amplitude \citep{Ertel2016}. Such variability can be interpreted as evidence for non-steady replenishment \citep[e.g., comet disruption,][]{2019AJ....157..202S}, dust distribution asymmetries \citep{2020A&A...635A..10S}, or, finally, the presence of a companion, whose impact varies with its orbital position. This last scenario in particular has been ruled out in a number of systems, based on Gaussian flux ratio estimations and near-zero ($\lesssim 1^{\circ}$) closure phase measurements, which suggests there are no point-like asymmetries \citep[e.g.,][]{2007A&A...475..243D,2009ApJ...704..150A,Kirchschlager2020}. However, more recent studies have revealed multiple planets in the inner regions of some of these systems \citep{2017A&A...605A.103F, 2017AJ....154..135F,2019A&A...623A..72K}, including a candidate around Vega \citep{2021AJ....161..157H}.

Despite these companions being likely too faint to cause the detected visibility deficits, this highlights our evolving understanding of these observations. In particular, it raises the important question of whether the near-zero closure phase method to rule out companions is truly reliable. In other words, we suggest that some of these infrared observations could be attributed to faint, falsely undetected companions rather than dust. To investigate this question, here we  present  an analysis of the conditions under which a companion might falsely
remain undetected due to near-zero closure phases, leading to the misinterpretation of observational data caused by the similarity of dust and
companion signatures in visibilities and closure phases, particularly when observations are limited in number.

One prominent example of a main sequences star with a detected visibility deficit is \(\kappa\) Tuc~A (HD 7788). It is an F6 IV-V star located in the constellation Tucana at a distance of \(20.99\) parsec \citep{2016A&A...595A...1G,2023A&A...674A...1G}. The star has an effective temperature of \(6474 \, \mathrm{K}\), a stellar mass of approximately \(1.35 \, {\rm M}_\odot\) \citep{Fuhrmann2017}, and an estimated age of \(2 \, \mathrm{Gyr}\) \citep{Tokovinin2020}. The variability of its potentially extended excess emission makes it an exceptional candidate for studying whether a stellar or planetary companion may cause the observed visibility deficit. The NIR observations in the H band ($\lambda\!\sim\!1.65 \, \mu\mathrm{m}$) with VLTI's Precision Integrated Optics Near
Infrared Experiment \citep[VLTI/PIONIER,][]{2011A&A...535A..67L} revealed a temporal flux variability, with a measured significant dust-to-star flux ratios of \(f = (1.43 \pm 0.17)\%\) in 2012 and \(f = (1.16 \pm 0.18)\%\) in 2014, but no significant excess (\(f = 0.07 \pm 0.16\%\)) detected in 2013 \citep{Ertel2014, Ertel2016}. Recent observations with the VLTI and the Multi AperTure mid-Infrared SpectroScopic Experiment instrument \citep[MATISSE,][]{2014Msngr.157....5L,2022A&A...659A.192L} have revealed the presence of significant visibility deficit in the L band in the MIR (\(\approx3.3\text{--}4 \, \mu\mathrm{m}\)) that was attributed to the thermal emission of hot exozodiacal dust with a dust-to-star flux ratio of \(5\text{--}7\%\) \citep{Kirchschlager2020}. The dust was modeled as being distributed in a narrow ring located at a distance of \(0.1\text{--}0.29 \, \mathrm{au}\) from the star, corresponding to a temperature range of \(940 \text{--} 1430 \, \mathrm{K}\). Despite these findings, to this day, the origin of the variability in the dust-to-star flux ratio remains unsolved and the existence of hot exozodiacal dust itself is still a puzzling matter.

In this study, we build on previous efforts \citep[e.g.,][]{2012A&A...541A..89L} and examine how closely the observed signatures that are commonly attributed to hot exozodiacal dust could be replicated by a faint companion, using $\kappa$~Tuc~A as an example. In Sect.~\ref{sec:methods}, we introduce our model, consisting of a limb-darkened central star and an off-axis companion. Section~\ref{sec:results} presents our results of the modeled signatures of a faint companion. This includes a derivation of upper limits of interferometric quantities, calculated non-detection probabilities, the role of configurations and instruments, and exemplarily an analysis of companions on two differently inclined orbits as well as a reevaluation of a MATISSE observation of $\kappa$~Tuc~A. We discuss these results in Sect.~\ref{sec:discussion}, where we reevaluate past companion rejection criteria and argue that the underlying assumptions may not have been universally valid. Furthermore, we propose an additional, novel method for distinguishing hot exozodiacal dust from faint companions. A summary of our findings together with a Jupyter notebook, published alongside this paper, are presented in Sect.~\ref{sec:summary}.

\section{Methods}
\label{sec:methods}
\begin{figure}
    \resizebox{\hsize}{!}{
        \includegraphics[clip]{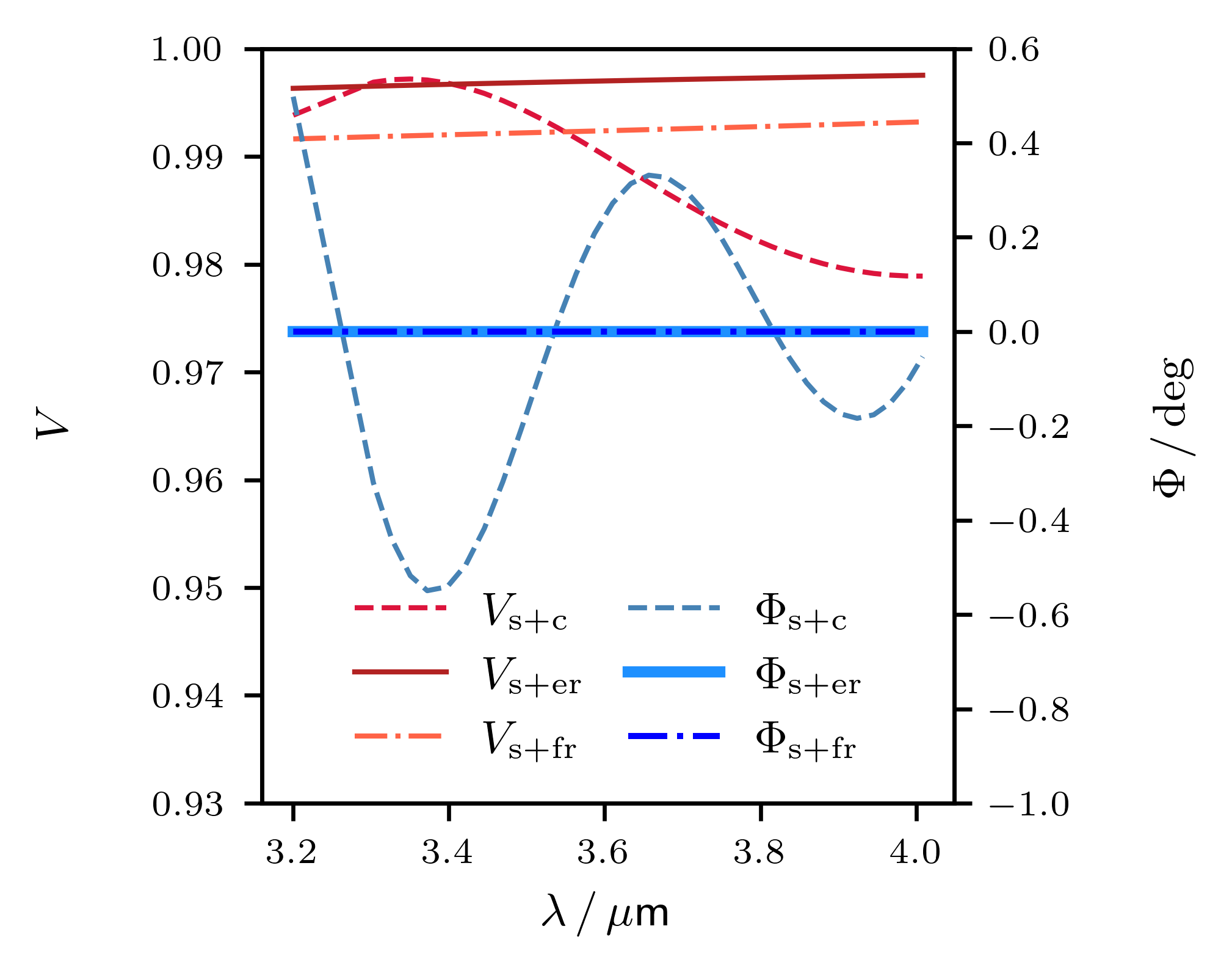}
    }
    \caption{Expected visibility amplitudes, $V$, and closure phases, $\Phi$, of a binary system composed of a central star and a companion (dashed lines, denoted by the index \texttt{s+c}), exemplarily modeled for $\kappa$~Tuc~A  and a faint companion at the projected distance of $\approx 1.4\,$au, assuming a companion-to-star flux ratio of $f=1\%$ and a fixed telescope triplet \citep{VLTI_Manual_2024}. In comparison to that, the plot shows the modeled visibilities and closure phases of an edge-on (solid lines, denoted by the index \texttt{s+er}) and face-on (dash-dot lines, denoted by the index \texttt{s+fr}) hot dust ring with the inner radius $R_{\rm in}=0.1\,$au \citep{Kirchschlager2020} and a flux ratio of $f\approx 1\%$.}
    \label{fig:cps_and_vis_example}
\end{figure}
In the following, we present the equations to describe the interferometric properties of a limb-darkened star and a companion in Sect. \ref{sec:vis_and_clos}. This is followed by a derivation of the impact a faint companion has on the measured visibilities and closure phases in Sect. \ref{sec:theoretical_limits}. Furthermore, we derive linear approximations of these equations in terms of the companion-to-star flux ratio, which provides the basis for the scalability of our results shown in later sections. In Sect. \ref{sec:instruments}, the assumed instruments are described, the properties of which are used in our numerical simulations, presented in Sect. \ref{sec:simulations}.

\subsection{Visibility and closure phases}
\label{sec:vis_and_clos}
In this study, we analyzed two key interferometric observables, the visibilities and closure phases, with regard to their potential to support or reject close-in faint companions in the analysis of infrared excess found for selected main sequence stars. Analyzing these features and comparing them to observable features of systems that include hot exozodiacal dust is important to identify key differences, which may allow us to distinguish these phenomena. The visibility amplitude of the partly resolved stellar photospheric emission with limb-darkening at an observing wavelength $\lambda$ is given by 
\begin{equation}
    \label{eq:star_vis}
    V_{\text{s}} = \frac{6}{3-u_{\lambda}}\left( (1-u_{\lambda})\frac{J_1(\beta)}{\beta}+u_{\lambda}\sqrt{\frac{\pi}{2}}\frac{J_{1.5}(\beta)}{\beta^{1.5}} \right),
\end{equation}
where $\beta=\pi\Theta_{\text{s}}B/\lambda$.
All stellar parameters were adopted for $\kappa$~Tuc~A. The angle diameter of the star is $\Theta_{\text{s}}=0.739\,$mas \citep{Ertel2014}, $u_{\lambda}$ is the limb-darkening coefficient \citep[0.25 for the H band, 0.22 for the K band, and 0.19 for the L band,][]{1995A&AS..114..247C,2011MNRAS.413.1515H}, $J_1(\beta)$ and $J_{1.5}(\beta)$ are Bessel functions of the first kind, and $B$ is the distance between two telescopes projected onto the sky, also called the baseline \citep{1974MNRAS.167..475H}. For each telescope pair with its corresponding baseline, a single visibility measurement is obtained.
The companion was modeled as a point source with a complex visibility of
\begin{equation}
    \mathcal{V}_{\text{c}} = \exp \left[ \mathrm{i}\psi_{\text{c}}\right] = \exp \left[-2\pi \mathrm{i}\left(\frac{u}{\lambda}x+\frac{v}{\lambda}y\right)\right],
    \label{eq:visib_comp}
\end{equation}
where $u$ and $v$ are the coordinates in the Fourier space, $x$ and $y$ are Cartesian coordinates, while $\mathcal{V}_{\text{c}}$ is a complex number with an absolute value, $V_{\text{c}} = \left\lvert \mathcal{V}_{\text{c}} \right\rvert = 1$, and the phase, $\psi_{\text{c}}$. The assumption of a point source is justified for two reasons. First, the distance of a potential companion from the star has a significantly greater effect on the visibility than the diameter of the companion.  Second, given the faintness of the companion, any additional effect from it being partially resolved is suppressed and expected to be negligible.

The total complex visibility of a system is obtained by adding and weighting the complex visibilities of all components of the system.
The measured value of the total visibility amplitude of a system that consists of a star and a companion with the flux of the star, $f_{\text{s}}$, and the flux of the companion, $f_{\text{c}}$, is then given by 
\begin{equation}
    V_{\text{s+c}} =\left\lvert\mathcal{V}_{\text{s+c}} \right\rvert=\left\lvert \frac{V_{\text{s}}f_{\text{s}}+\mathcal{V}_{\text{c}}f_{\text{c}}}{f_{\text{s}}+f_{\text{c}}}\right\rvert=\left\lvert \frac{V_{\text{s}}+\mathcal{V}_{\text{c}}f}{1+f}\right\rvert,
    \label{eq:v_ss_eq_A}
\end{equation}
where $f=f_{\text{c}}/f_{\text{s}}$ \citep{2007A&A...475..243D}. When performing an interferometric measurement, the properties of the entire system are accessed. Consequently, the visibility deficit induced by an existing companion would be given by the difference between the stellar visibility amplitude and the absolute value of the visibility amplitude of the combined system, expressed as
\begin{equation}
    \Delta V =  V_{\text{s}}- V_{\text{s+c}}.
    \label{eq:dv}
\end{equation}
Hereafter, we refer to visibility amplitudes simply as visibilities.
Another important interferometric quantity is the closure phase, $\Phi$, which is measured across a closed triangle of baselines. When combining three or more telescopes, each telescope triplet provides such a closure phase measurement, expressed as
\begin{equation}
    \Phi = \varphi_{12}+\varphi_{23} +\varphi_{31},
    \label{eq:closure_phases_general_eq}
\end{equation}
where 
\begin{equation}
    \varphi_{ij}=\arctan\left(\mathfrak{Im}(\mathcal{V}_{ij})/\mathfrak{Re}(\mathcal{V}_{ij})\right)
    \label{eq:phase}
\end{equation}
 is the phase of a complex visibility, $\mathcal{V}_{ij}$, measured between the $i$th and $j$th telescopes. The closure phase is sensitive to deviations from point symmetry and is essential for identifying systems such as binaries. These systems differ significantly from point-symmetric dust distributions as well as from the rings, both face-on and edge-on, of hot exozodiacal dust, as the point-asymmetry in the case of the binary is stronger assuming the same flux ratio. Such dust-containing systems can be expected to have closure phases close to zero, as demonstrated in Fig.~\ref{fig:cps_and_vis_example}. This plot shows exemplarily the modeled visibilities and closure phases of a system composed of a limb-darkened star and a companion as a function of the wavelength in the L band (dashed lines). Here, the non-zero closure phase is a result of the asymmetry induced by the additional presence of the companion. In contrast, the solid lines show the visibilities and closure phases of a narrow hot dust ring at $R_{\rm in}$, which was exemplarily modeled with DMS \citep{2018A&A...618A..38K,2020PhDT........31K}. The visibilities and phases were calculated using the computational library galario \citep{2018MNRAS.476.4527T}. While both scenarios can reproduce similar levels of infrared excess, their interferometric signatures differ. These differences provide a basis for distinguishing between these two phenomena. However, there are configurations (e.g., small angular separations) where their signatures might overlap. A careful analysis of the wavelength dependence and baseline configuration is therefore necessary to disentangle the two scenarios reliably.\\

\subsection{Signature of a faint companion}
\label{sec:theoretical_limits}
In this subsection, we study the maximum impact a faint companion can have on interferometric measurements. For this purpose, we derived the maximum visibility deficits and closure phases that can be induced by a single companion, building on the work of \citet{2012A&A...541A..89L}. 
To analyze the dependence of the total visibility amplitude of the system \( V_{\text{s+c}}(f) \) on the flux of the faint companion (i.e., $f=f_{\text{c}}/f_{\text{s}}\ll 1$), we can expand Eq.~\eqref{eq:v_ss_eq_A} as a Taylor series to the first order in the flux ratio (for details, see Fig.~\ref{app:deriv} in the appendix). Later on, this linearity  allows us to derive results that are scalable with regard to the flux ratio, which is required for the generality of our results and allows for easy applicability. For the following derivations, we additionally assume that the stellar visibility (Eq. \ref{eq:star_vis}) is close to one, which is generally the case for all wavelengths considered in this study. This yields the linear approximations,
\begin{equation}
  \mathcal{V}_{\text{s+c}}(f)=V_{\text{s}}+(\mathcal{V}_{\text{c}}-V_{\text{s}})f
  \label{eq:taylor_sys_complex}
\end{equation}
and
\begin{equation}
  V_{\text{s+c}}(f)=V_{\text{s}}+(\mathfrak{Re}\left\{\mathcal{V}_{\text{c}}\right\}-V_{\text{s}})f
  \label{eq:taylor_sys}
,\end{equation}
with a visibility deficit, according to Eq.~\eqref{eq:dv}, expressed as
\begin{equation}
  \Delta V(f)=(V_{\text{s}}-\mathfrak{Re}\left\{{\mathcal{V}_{\text{c}}}\right\})f.
  \label{eq:dv_f}
\end{equation}
With the maximum visibility of a star being one and the minimum real part of a visibility of a companion being equal to minus one, we can derive an upper limit for the observed visibility deficit induced by a faint companion, which is linear in its flux ratio, $f$, and given by
\begin{equation}
  \left\lvert\Delta V(f)\right\lvert\leq2 f.
  \label{eq:max_delta_v}
\end{equation}
Consequently, a faint companion can cause a visibility deficit up to twice its percentage flux contribution for $f\ll1$. Moreover, we are now able to put a lower limit on the flux of a companion, based on the visibility deficit measurement, namely, $f_{\text{min}}=\left\lvert\Delta V(f)\right\lvert/2$. 

Following the same principle, we were also able to restrict the maximum of the closure phase dependent on the flux ratio, $f$. The combination of three telescopes as a closed triangle gives three values of the visibility. According to Eq.~\eqref{eq:taylor_sys_complex}, the visibility of a pair of telescopes $i$ and $j$ can be written as
\begin{equation}
    \mathcal{V}_{\text{s+c},ij}(f)=V_{\text{s},ij}+(\mathcal{V}_{\text{c},ij}-V_{\text{s},ij})f, \\
  \label{eq:re_im}
\end{equation}
where $\mathcal{V}_{\text{c},ij} = \mathrm{e}^{\mathrm{i}\psi_{ij}}$ (see Eq. \ref{eq:visib_comp}). To the first order from Eq.~\eqref{eq:phase}, the corresponding phase $\varphi_{ij}$ of the whole system in radian then equals 
\begin{equation}
    \varphi_{ij} = \arctan \left(\frac{f\sin \psi_{ij}}{V_{\text{s},ij} + (\cos \psi_{ij} - V_{\text{s},ij})f}\right).
\end{equation}
Given the small flux ratio, this expression can be approximated, to the first order in terms of $f,$ as
\begin{equation}
    \varphi_{ij} = \frac{f\sin \psi_{ij}}{V_{\text{s},ij}}.
\end{equation}
With that, the closure phase is equal to
\begin{equation}
    \Phi=f\left(\frac{\sin \psi_{12}}{V_{\text{s,12}}}+\frac{\sin \psi_{23}}{V_{\text{s,23}}}+\frac{\sin \psi_{31}}{V_{\text{s,31}}}\right),
    \label{eq:CP}
\end{equation}
which gives rise to
\begin{equation}
    \left\vert \Phi\right\vert \leq f\left(\frac{1}{V_{\text{s,12}}}+\frac{1}{V_{\text{s,23}}}+\frac{1}{V_{\text{s,31}}}\right)
\end{equation}
and, therefore,
\begin{equation}
    \left\vert \Phi \right\vert \leq \frac{3f}{\min\limits_{i,j \in \{1,2,3\}} \left\{V_{\text{s},ij}\right\}}.
    \label{eq:max_cp}
\end{equation}
Equations \eqref{eq:dv_f} and \eqref{eq:CP} indicate that the induced visibility deficit and the closure phases exhibit a linear dependence on the flux ratio, \(f\), for faint companions. Furthermore, we validated these calculations through numerical tests and found that the linear relationship holds reasonably well for flux ratios, $f_{\text{c}}/f_{\text{s}}$, up to $\approx$10\%.
\subsection{Instruments}
\label{sec:instruments}
We modeled the observed quantities for the VLTI for three state-of-the-art instruments: PIONIER, GRAVITY, and MATISSE. These instruments offer complementary capabilities across different wavelength bands, which is required for distinguishing different sources and therefore enables a comprehensive study of circumstellar environments. Their capability to combine the light from different telescopes (in this case: four) allows for high-resolution measurements of the inner-most circumstellar objects and structures.These three instruments can measure both visibilities and closure phases, enabling the detection of faint circumstellar features with high contrast.
PIONIER operates in the H band (\(\sim1.65 \, \mu\mathrm{m}\)), providing high-precision visibility measurements that are useful for probing NIR excess emission.

The instrument GRAVITY \citep{2017A&A...602A..94G} is an interferometer working in the K band (\(\sim2.2 \, \mu\mathrm{m}\)). It is particularly well-suited for studying the intermediate wavelengths between the NIR and MIR domains. The MATISSE instrument extends the wavelength coverage to the L, M, and N bands ($\sim3.4$--$13\, \mu\mathrm{m}$) and has been shown to be capable of determining the properties of the dust more precisely than is possible using H and K band measurements \citep{2018MNRAS.473.2633K}. In this study, we simulated only the L band wavelengths of MATISSE representative for the nearest hot environment of the star.

Moreover, the VLTI supports observations with four different Auxiliary Telescope (AT) configurations \citep{VLTI_Manual_2024}: small (A0-B2-D0-C1), medium (K0-G2-D0-J3), large (A0-G1-J2-K0), and extended (A0-B5-J2-J6). These configurations determine the baseline lengths between the telescopes and thus the spatial resolution of the observations. For small, large, and extended configurations, we selected six representative projected baselines chosen with Aspro2 \citep{10.1117/12.2234426} for $\kappa$~Tuc~A. The medium configuration was taken from the observations of \citet{Kirchschlager2020}. This number matches the VLTI capability of combining four ATs and results in six baselines. The lengths of the baselines for the small, medium, large, and extended configuration are between 10--31\,m, 35--96\,m, 35--129\,m, and 37--153\,m, respectively, corresponding to a minimum resolution of about 82.5\,mas (i.e., 1.73\,au) in the L band (10\,m baseline, 4$\,\mu$m), and a maximum resolution of about 2.06\,mas (i.e., 0.043\,au) in the H band (150\,m baseline, 1.5$\,\mu$m). For each configuration and each VLTI instrument we calculated and analyzed closure phases and visibility deficits of the system with a central star and a faint close-in companion at different projected positions, using the analytical equations presented in this section. Additionally, we consider the combination of different instruments and configurations, as it enables multi-wavelength, high-resolution observations, which can aid in the characterization of the inner environment of the star and its potential origin of variability. We note that we have made a Jupyter notebook publicly available alongside this paper, which gives similar calculations to those presented throughout this paper, with the goal of helping researchers evaluate existing observations and plan future ones.\footnote{\url{https://github.com/kate-tis/non_detection_of_faint_companions}.}

\begin{figure*}
    \resizebox{\hsize}{!}{
        \includegraphics[clip]{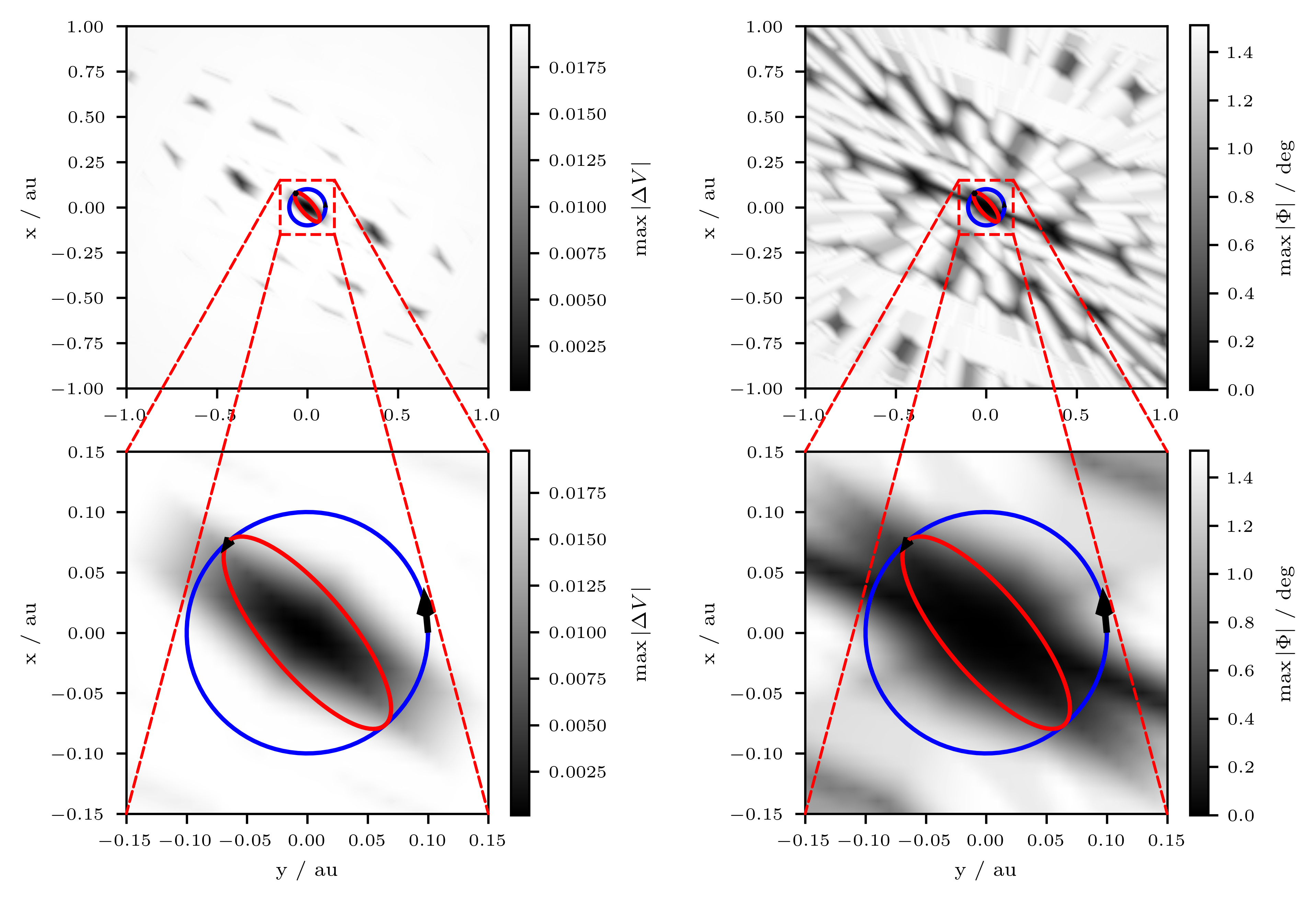}
    }
    \caption{Maximum visibility deficits (left) and closure phases (right) of the system in dependence on the position of the assumed companion for all simulated MATISSE wavelengths and medium configuration. Field of view of \( 2 \, \mathrm{au} \times 2 \, \mathrm{au}  \), or \( 0.09 \, \mathrm{as} \times 0.09 \, \mathrm{as}\),  (top) and \( 0.3 \, \mathrm{au} \times 0.3 \, \mathrm{au}  \), or \( 0.014 \, \mathrm{as} \times 0.014 \, \mathrm{as}\), (bottom). The arrows show the direction of the orbiting companion. The blue and red solid lines show the theoretical orbits of a companion,  explained in Sect.~\ref{sec:blind_spots}.}
    \label{fig:max_dvs}
\end{figure*}

\subsection{Numerical simulations}
\label{sec:simulations}
Following the general derivations described in Sect.~\ref{sec:theoretical_limits},  for our simulations, we used the analytically calculated visibility deficits and closure phases for a binary system consisting of a main-sequence central star (here, $\kappa$~Tuc~A) and a faint companion at different locations within a field of view of \( 2 \, \mathrm{au} \times 2 \, \mathrm{au}  \), corresponding to an angular separation of up to \( 0.07 \)\,as from the star. For that purpose, we applied Eqs.~\eqref{eq:v_ss_eq_A} and~\eqref{eq:closure_phases_general_eq} using 33 linearly sampled wavelengths for each considered instrument (PIONIER, GRAVITY, and MATISSE in the H, K, and L bands, respectively), and configuration (small, medium, large, and extended). We note that the number of wavelengths were chosen arbitrarily to find the strongest possible signal and do not correspond to the spectral coverage of the instruments. Since the closure phase is strongly dependent on the wavelength (see Fig.~\ref{fig:cps_and_vis_example}), it is crucial to consider multiple wavelengths to fully capture this dependency to avoid any bias that could be introduced by representing a band by a single wavelength. The maximum and minimum simulated wavelengths for each band are shown in Table~\ref{tab:max_min_lambda}.
\begin{table}
\caption{Maximum and minimum simulated observational wavelengths for each VLTI instrument.}
\centering
\begin{tabular}{c|c|c}
\hline\hline
 & $\lambda_{\text{min}}\,/\,\mu$m & $\lambda_{\text{max}}\,/\,\mu$m \\
\hline
PIONIER (H band) & 1.5& 1.8 \\
GRAVITY (K band)  & 2.0 & 2.5 \\
MATISSE (L band) & 3.2 & 4.0 \\
\hline
\end{tabular}
\label{tab:max_min_lambda}
\end{table}
Then, for any combination of instrument and configuration, the maximum visibility deficit and closure phases were determined among all corresponding baselines and wavelengths. The maxima are particularly significant, as they represent the strongest effect the companion would exhibit at each projected position relative to the star. Therefore, they are crucial for assessing the feasibility to measure its impact and detect a companion. We assumed a flux ratio between the companion and the star of \( f = 1\% \). The companion was sequentially placed at each pixel position within the 201x201 pixel map, with a pixel size of $\approx0.01\,$au, and the six corresponding visibility deficits and four closure phases were calculated for each considered wavelength. Their absolute maximum was taken and is represented with color-coding. Although we did not vary the flux ratio, according to our previous results (see Eqs.~\ref{eq:dv} and~\ref{eq:CP}), both determined interferometric quantities can be scaled linearly with respect to $f$ for $f\lesssim10\%$. We also accounted for the sensitivity of PIONIER, GRAVITY, and MATISSE, which exhibit a dependence on the off-axis position of the companion in the field of view by reducing its flux according to a Gaussian profile. The full width at half maximum (FWHM) values used were 0.22\, as, 0.3\,as, and 0.6\,as, respectively. 
\section{Results}
\label{sec:results}
\subsection{Blind spots in visibility deficits and closure phases}
\label{sec:blind_spots}
For demonstration purposes, Fig.~\ref{fig:max_dvs} 
shows the results obtained for the MATISSE instrument in its medium configuration. The results for the PIONIER and GRAVITY medium configuration can be seen in Appendix~\ref{sec:pionier_gravity}. The upper plots show the region $\in\left[-1,1\right]\,$au and the lower plots show the zoomed-in version for the region $\in\left[-0.15,0.15\right]\,$au. The resulting patterns exhibit intricate structures due to the combination of different orientations and lengths of the six baselines. We note that the innermost dark region in these maps approximately corresponds to the inner working angle of the instrument ($\sim$ 6\,mas for MATISSE), within which the emission of the star can hardly be distinguished from the emission of the companion. The intricate patterns of visibilities and closure phases generally differ, as visibilities involve combinations of two telescopes, while closure phases are derived from the combination of three telescopes. The many dark spots on these maps show the projected location of a companion at which the visibility deficit respectively the closure phase would be equal or near to zero. Consequently, requiring a measured non-zero closure phase as a mandatory detection criterion for a potential companion could result in false negatives. Specifically, if a companion were located in one of these ``blind spots'' of the closure phase maps, it would remain undetected, even though its induced visibility deficit might still be measurable. This would result in the incorrect rejection of the companion as the source of the measured visibility deficit and lead to the erroneous conclusion that hot dust is present. Furthermore, we find that the projected regions of these blind spots in the visibility deficit maps (left) of Fig.~\ref{fig:max_dvs} are smaller than those in the closure phase maps (right). 
This indicates a greater chance of misinterpreting a companion as hot exozodiacal dust. This means that the companion would cause a visibility deficit, but  possibly without leaving a sufficiently significant impact on the closure phases. Lastly, we find that the maxima of the visibility deficits and closure phases found in these plots are close to the derived upper limits in Eqs.~\eqref{eq:max_delta_v} and~\eqref{eq:max_cp}. 

To illustrate the possibility that a companion can simultaneously cause a near-zero closure phase signal and significant visibility deficit, we selected two orbits (represented in the plots of Fig.~\ref{fig:max_dvs}
by red and blue ellipses) to examine the temporal variations of the maximum absolute visibility deficit, \( \left\vert \Delta V \right\vert \), and the maximum absolute closure phase, \( \left\vert \Phi \right\vert \), along these trajectories. Both orbits are intrinsically circular and have a radius of 0.1\,au, which is the radius where hot exozodiacal dust is assumed to be found for $\kappa$~Tuc~A, but they differ in their inclination and position angle. We note that these orbits were taken as an example and they do not correspond to real, physical orbits. The companion following the blue orbit would produce detectable signatures at certain points along its trajectory; whereas the companion on the red orbit would remain well inside the blind spot of the closure phase map. Thus, the latter is much more difficult to detect using this particular configuration.

\begin{figure*}
    \resizebox{\hsize}{!}{
        \includegraphics[clip]{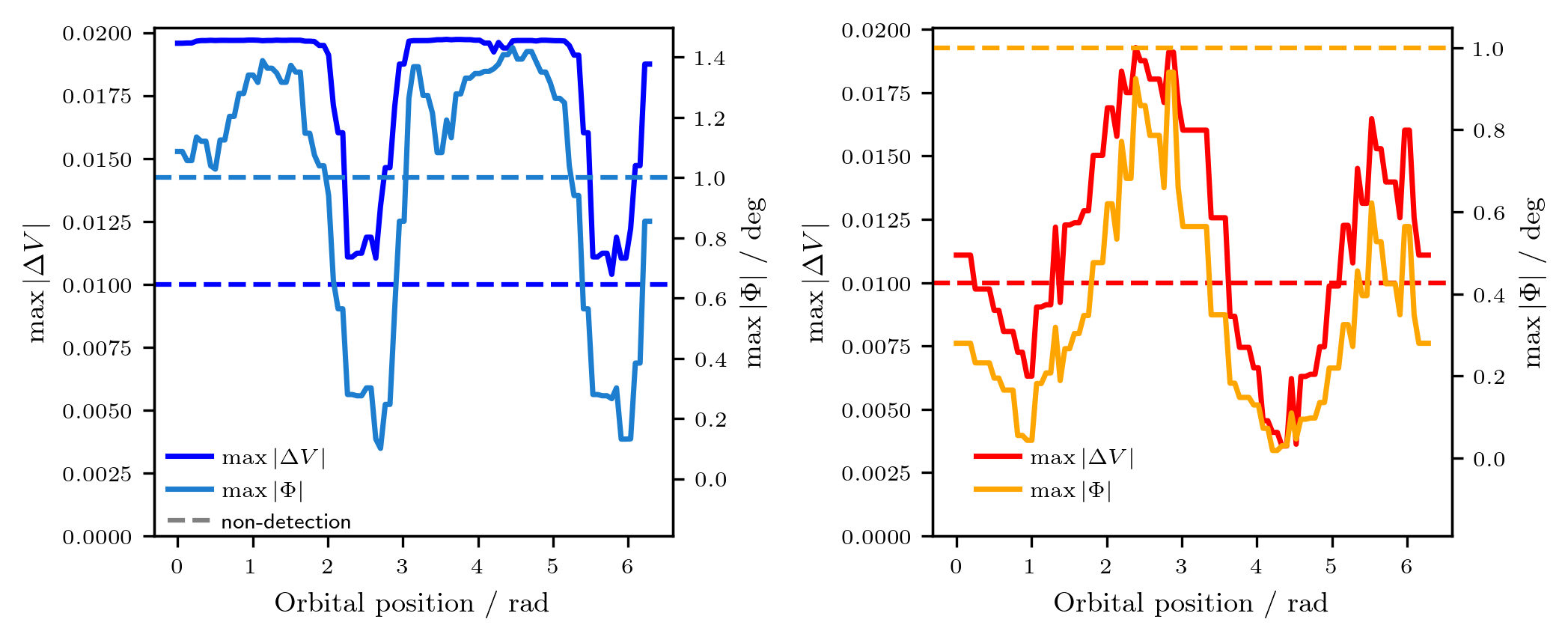}
    }
    \caption{Theoretical variation of the maximum visibility deficit $\max\left\lvert\Delta V \right\rvert$ and the maximum closure phase $\max\left\lvert\Phi \right\rvert$ along the trajectory of the face-on (blue) orbit (left) and inclined (red) orbit (right) in the direction shown by the arrows displayed in Fig. \ref{fig:max_dvs}.
    The dashed lines indicate the adopted generic detection limits: a visibility deficit of $\Delta V = 0.01$ and a closure phase of $\Phi = 1^{\circ}$. Their colors match those of the corresponding observables. Detection limits are discussed in detail in Sect.~\ref{sec:non_detection}.}
    \label{fig:blue_orb}
\end{figure*}

Figure \ref{fig:blue_orb} shows the changes in the maximum visibility deficit and the closure phase along the blue (left) and red (right) orbits and demonstrates, depending on the orbital position, that non-zero visibility deficits can be caused by a companion even when the closure phase remains close to zero. In case of this blue orbit, the maximum visibility deficit in its minimum would reach 0.01, while the maximum closure phase, which shows much stronger variability, would reach values of $<0.1^{\circ}$. In case of the red orbit. the companion would always stay at the threshold of $1^{\circ}$, even though it would still induce the visibility deficit of $\geq0.01$, which we consider a significant detection of visibility deficit. The dips in closure phase and visibility deficit curves correspond to the companion moving through blind spots (i.e., darker regions) in Fig.~\ref{fig:max_dvs}. 
This demonstrates, that under such conditions, the presence of a companion cannot be ruled out solely on the basis of closure phase measurements.  

\subsection{Non-detection probabilities in closure phases}
\label{sec:non_detection}
\begin{figure}
    \resizebox{\hsize}{!}{
        \includegraphics[clip]{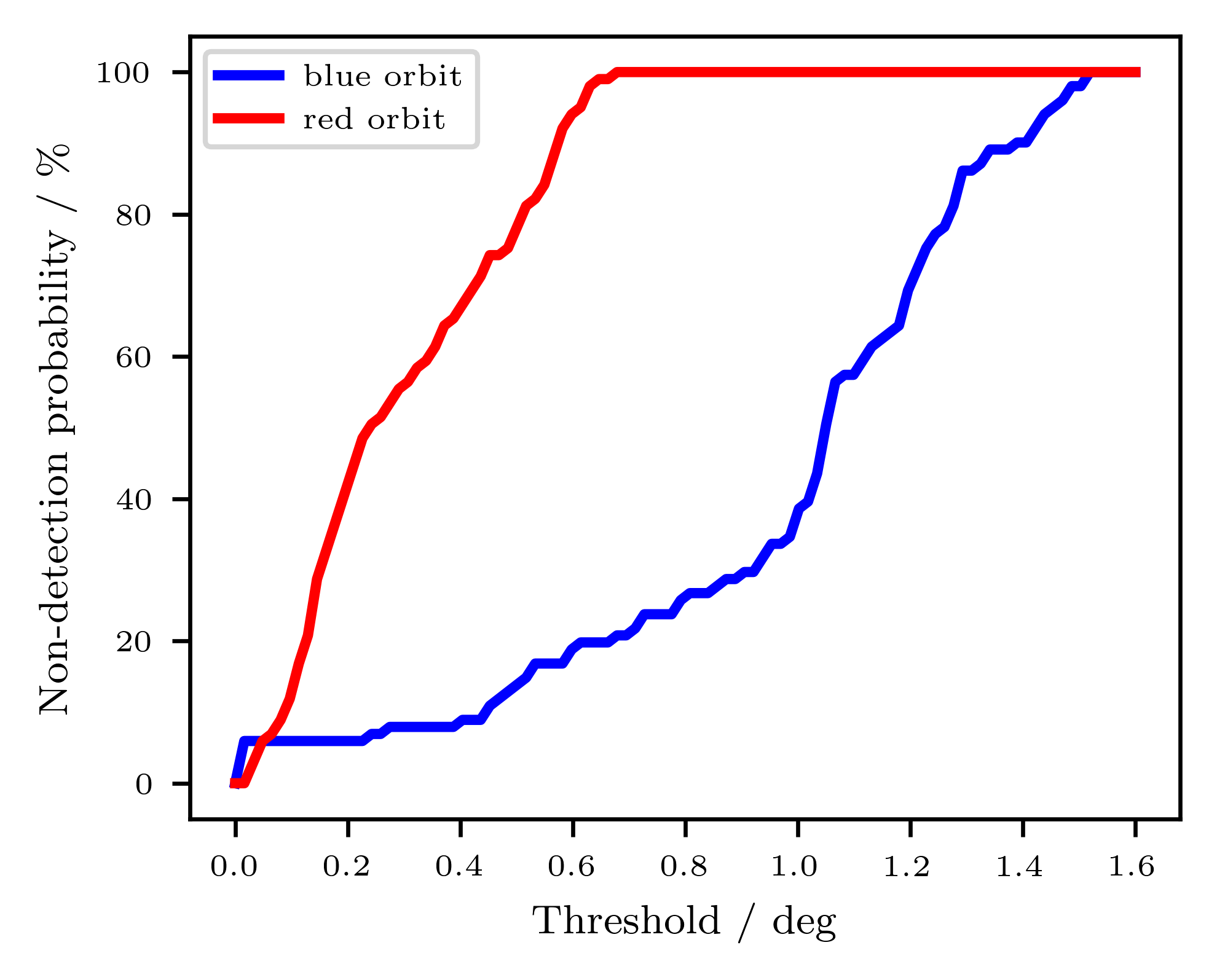}
    }
    \caption{Non-detection probabilities in closure phases as a function of the threshold for the face-on (blue) and inclined (red) orbits shown in Fig.~\ref{fig:max_dvs}. 
    For details, see Sect.~\ref{sec:theoretical_limits}.}
    \label{fig:probs_orbits}
\end{figure}

In this section, we assess the inherent non-detection probability of a faint companion for one observation based on the accuracy of closure phase measurements for the MATISSE medium configuration. We defined a non-detection as a measurement with a maximum closure phase below some threshold (i.e., the instrumental accuracy), $T$. Figure \ref{fig:probs_orbits} shows exemplarily the dependence of the non-detection probability on the threshold for two orbits assuming a flux ratio of $f=1\,\%$. The red (blue) curve corresponds to the inclined red (face-on blue) orbit from the closure maps (right) of Fig.~\ref{fig:max_dvs}. The probabilities were calculated by determining the closure phases of 101 equally spaced orbital positions. For a  threshold of $T\geq0.2^{\circ}$, which is the approximate error for closure phases in the L band, \citep[compare with Fig.~1 in][]{Kirchschlager2020}, the chosen red orbit, which has projected separations $\leq1\,$au, has a non-detection probability that is greater than $\sim$40\,\%. The blue face-on orbit has a probability that is greater than $\sim$10\,\%, making a successful detection in this case more likely. 

\begin{figure}
    \resizebox{\hsize}{!}{
        \includegraphics[clip]{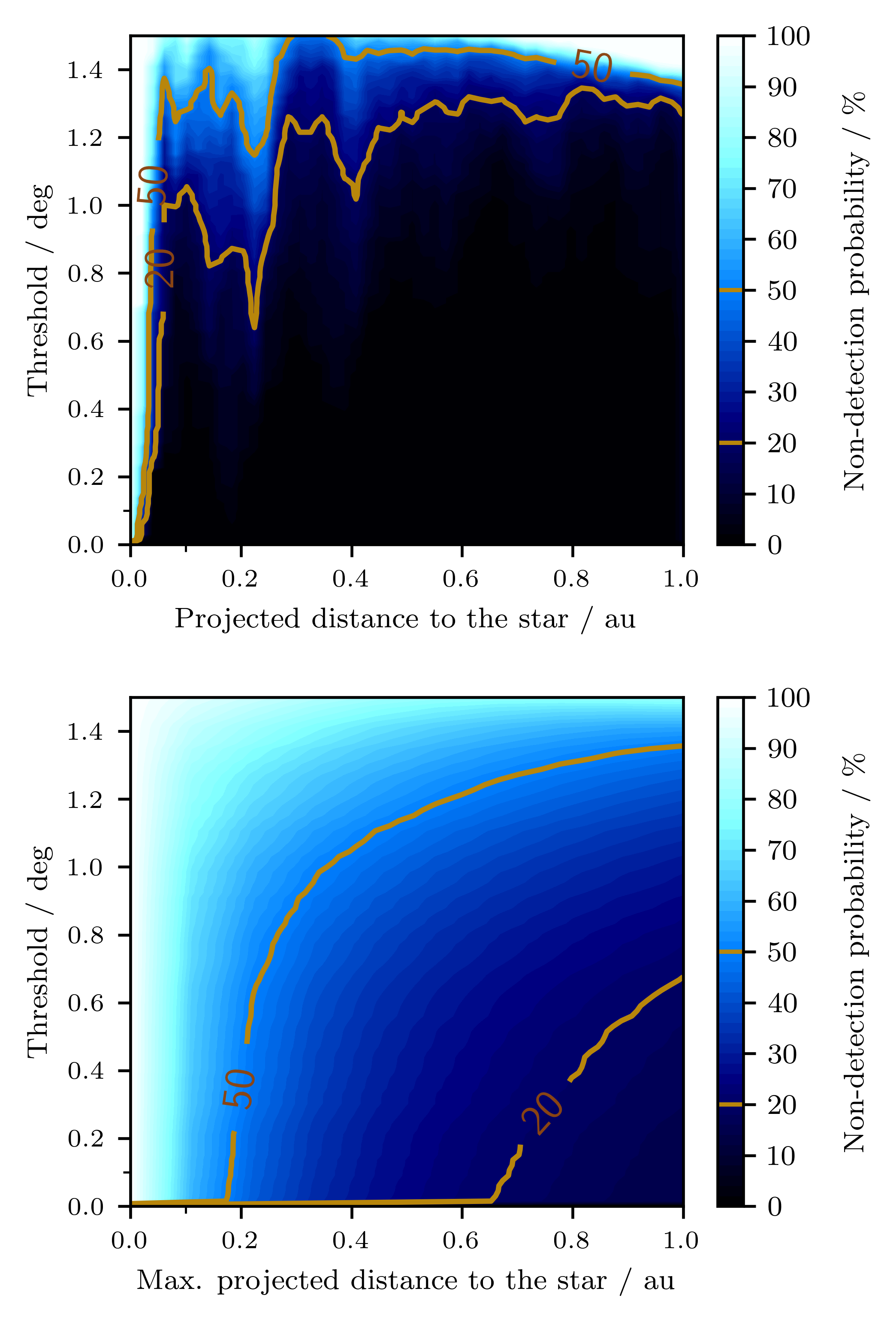}
    }
    \caption{Non-detection probabilities in closure phases for a companion at (top) and within (bottom) a given projected distance to the star in dependency of an observational accuracy (threshold).}
    \label{fig:probab_two}
\end{figure}

We note that although the determined probabilities cannot simply be scaled with the flux ratio, they remain unchanged for a constant value of the ratio between the threshold and the flux ratio, $T/f$, owing to the linearity of Eqs.~\eqref{eq:dv_f} and~\eqref{eq:CP}. This means that the non-detection probability of an object with $f=1\,\%$ and a threshold of $T=0.1^{\circ}$ are equal to the non-detection probabilities of a companion with $f=2\,\%$ and a threshold of $T=0.2^{\circ}$. In that regard, our results can be applied to any threshold value, which generally depends on the instrument, observing wavelength, atmospheric conditions, and observing time. 
Generally, the exact orbit of a companion and its position on the orbit can be assumed to be unknown. Figure \ref{fig:probab_two} therefore shows the calculated non-detection probabilities as a function of the projected distance and threshold. The upper plot illustrates the probability of failing to detect a companion located at a specific projected distance from the host star. In contrast, the bottom plot shows the non-detection probability for companions located within the specified projected distance, quantifying the likelihood of companions remaining undetected across different separations. To derive the non-detection probability for the upper plot, a companion has been placed consecutively at 101 equally spaced points at a fixed radius, for which the maximum closure phases were calculated and compared to the assumed threshold. If the threshold exceeded that value, it would count as a non-detection. The non-detection probability is then estimated by dividing the number of counted non-detections by the total number of considered companion placements. For the bottom plot, a similar method for deriving the non-detection probability has been applied. However, in this case the companion was placed on a cartesian grid using 201$\times$201 pixels covering a width of 2\,au$\times$2\,au. These plots show that the non-detection probability reaches values close to 100\,\% for separations below $\sim0.05\,$au and thresholds above $0.1^{\circ}$, which is linked to the inner working angle of the interferometer. Outside that radius, the probability significantly decreases for the displayed range of fixed threshold values. Furthermore, brighter vertical stripes in the upper plot indicate that the probability is not a monotone function of the separation. These regions of higher probability originate from a higher abundance of blind spots found at certain projected distances in the intricate pattern of the closure phase maps (right) shown in Fig.~\ref{fig:max_dvs}. The bottom plot of Fig.~\ref{fig:probab_two} additionally shows that the non-detection probability decreases if an assumed companion can be located at higher projected separations from the star, as long as the detection efficiency of the instrument is sufficiently high. The highest possible threshold reached in both figures corresponds to $\approx 1.5^{\circ}$ which is in agreement with the maximum closure phase estimation from Eq.~\eqref{eq:max_cp}. For a threshold higher than this value, the non-detection probability equals 100\%; this is because, with the given flux ratio, the system cannot reach higher closure phases.

\begin{figure}
    \resizebox{\hsize}{!}{
        \includegraphics[clip]{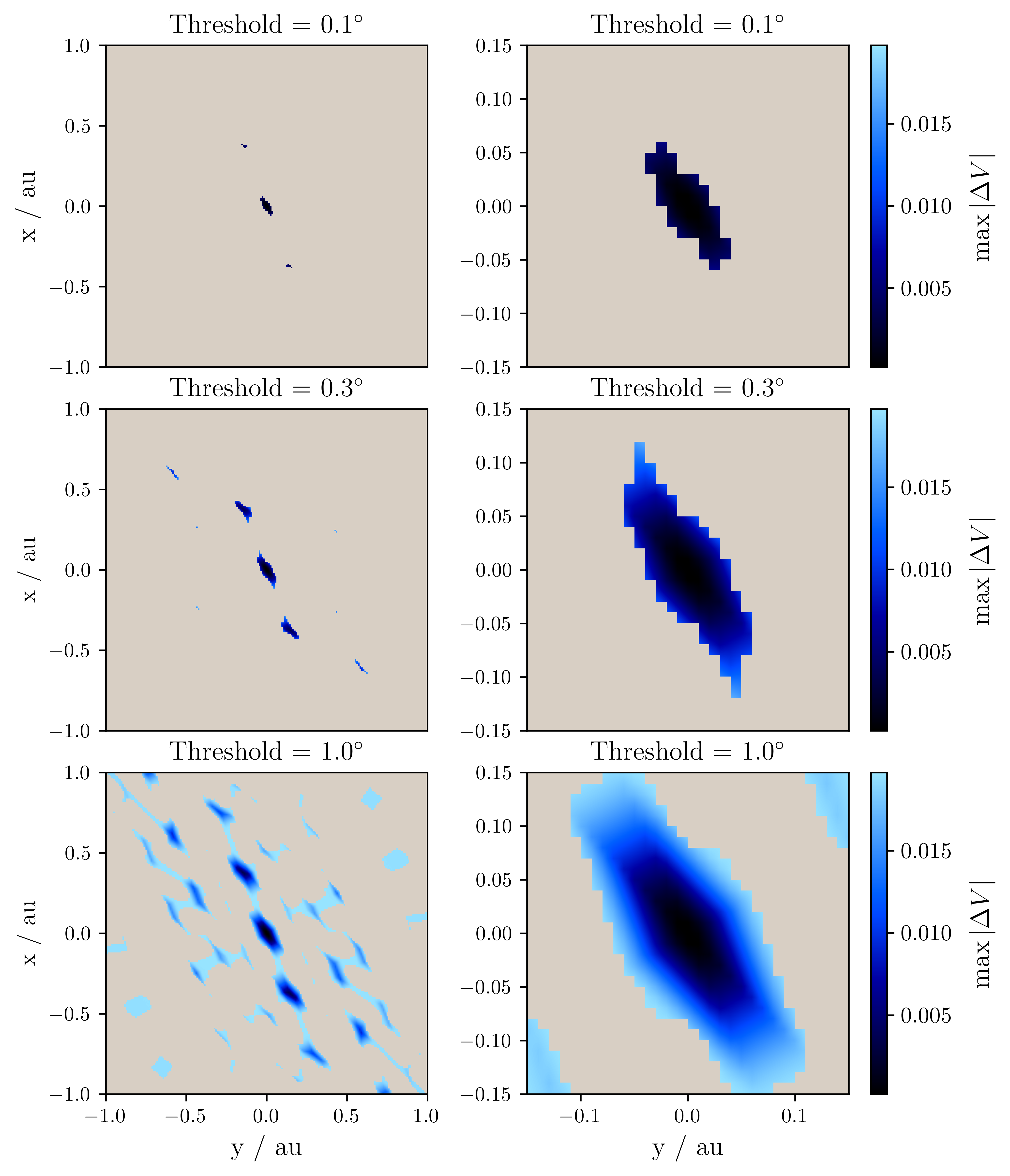}
    }
    \caption{Left: Maximum visibility deficit $\max\left|\Delta V\right|$ within regions where the closure phase satisfies $\left|\Phi\right| \leq T$ (non-detection) for the medium MATISSE configuration and the flux ratio $f=1\%$. The gray areas show the regions where $\left|\Phi\right| > T$ (detection). Right: Zoom-in with the field of view of \( 0.3 \, \mathrm{au} \times 0.3 \, \mathrm{au}  \), or \( 0.014 \, \mathrm{as} \times 0.014 \, \mathrm{as}\).}
    \label{fig:dvs_with_mask}
\end{figure}

Assuming non-detections in closure phases at thresholds of $T = 0.1^\circ$, $0.3^\circ$, and $1^\circ$, we calculated the highest visibility deficits for companions residing within the projected regions of the blind spots. These visibility deficits are crucial as they can help us distinguish between hot exozodiacal dust and a companion as the origin of a detected deficit. These regions are shown in Fig.~\ref{fig:dvs_with_mask} in blue for the medium configuration of MATISSE. In contrast, the gray areas indicate regions where closure phases exceed the given threshold, implying that a faint companion would be detectable in those areas. The plots on the left display a field of view of \( 2 \, \mathrm{au} \times 2 \, \mathrm{au}  \), while the plots on the right show a zoomed-in region within \( 0.3 \, \mathrm{au} \times 0.3 \, \mathrm{au}  \).

Based on these results, visibility deficits of up to $1.3f$ can occur even when closure phases are below $T = 0.1^\circ$, depending on the position of the faint companion. However, as shown in Fig.~\ref{fig:probab_two}, the probability of not detecting a companion within 1\,au is below 20\,\% for such a small threshold for this instrument and configuration. This suggests that in the case of a non-detection of an existing companion with this level of detection accuracy, it is more probable that a companion as origin of the deficit would be located closer to the star.

We performed the same analysis for all instruments and configurations for the three assumed detection thresholds  of $T=0.1^{\circ}$, $0.3^{\circ}$, and $1^{\circ}$. We calculated the non-detection probabilities for a potential companion anywhere within three projected separations of $r_{\text{max}} = 0.1$, $0.5$, and $1\,$au. The results of these calculations are summarized in Table \ref{tab:single_inst}. Additionally, we determined the highest visibility deficit corresponding to each combination of instrument, configuration, threshold and projected separation, as summarized in Table~\ref{tab:single_dvmax}. 
We find that regardless of the chosen instrument, the small configurations will fail to detect the faint companion in the closure phases within 0.1\,au with a probability of $\geq$75\%, given a 0.1$^{\circ}$ threshold. Still, in case of a non-detection, the highest possible induced visibility deficit is $\left|\Delta V\right|=0.9f=0.009$. For a threshold of 1$^{\circ}$ this probability reaches even 97-100\%. Nonetheless, the corresponding visibility deficit can nearly reach the theoretical upper limit of $0.02$ in case of a non-detection. Moreover, with a detection threshold of 0.1$^{\circ}$, we will not be able to detect a companion by analyzing the closure phases with a probability of $\geq$ 59\% if its projected separation is $\leq$0.1\,au, with a corresponding highest $\left|\Delta V\right|$ value of $0.02$. 

Our results in Table~\ref{tab:single_inst} suggest that the lowest non-detection probabilities correspond to the larger configurations (large and extended), assuming a fixed ratio of threshold to flux ratio. Moreover, we find that the non-detection probability decreases with decreasing observing wavelength, assuming a fixed $T/f$ ratio, suggesting higher detection probabilities with PIONIER compared to GRAVITY and MATISSE. These results are largely explained by the reduced size of blind spots at shorter observing wavelength and larger baselines, which is linked to the resolution of the interferometer. Based on the results in Table~\ref{tab:single_dvmax}, we find an intricate wavelength-dependence of the highest visibility deficit with a dependence on the position of the companion in the case of a non-detection. Furthermore, for all instruments, the highest visibility deficit for the large and extended configurations is lower than that of the small and medium configuration. 
In Sect.~\ref{sec:discussion}, its implications will be further discussed and used as a basis for describing a structured approach for distinguishing hot exozodiacal dust from faint companions. 
\begin{table*}
\caption{
Non-detection probabilities for each instrument and configuration.
}     
\centering          
\begin{tabular}{c c|c c c|c c c|c c c}     
\hline\hline       

 &  & \multicolumn{3}{c|}{$T=0.1^\circ$} & \multicolumn{3}{c|}{$T=0.3^\circ$} & \multicolumn{3}{c}{$T=1^\circ$} \\ 
\cline{3-11}
 \multicolumn{2}{c|}{Instrument \&} & \multicolumn{3}{c|}{$r_{\text{max}}$ / au} &\multicolumn{3}{c|}{$r_{\text{max}}$ / au}&\multicolumn{3}{c}{$r_{\text{max}}$ / au}  \\
\multicolumn{2}{c|}{Configuration}  & $0.1$ & $0.5$ & $1.0$ 
   & $0.1$ & $0.5$ & $1.0$ 
   & $0.1$ & $0.5$ & $1.0$ \\ 
\hline                    
   \multirow{4}{*}{MATISSE} & s & 95 & 51 & 34 & 100 & 64 & 44 & 100 & 80 & 64 \\  
    &m & 67 & 27    & 17 & 75 & 33 & 21 & 87 & 50 & 38 \\
   &l & 63 & 24     & 15 & 68 & 28 & 18 & 80 & 46 & 34 \\
   &e & 61 & 24    & 15 & 65 & 26 & 16 & 76 & 42 & 29 \\
   \hline
  \multirow{4}{*}{GRAVITY} & s & 82& 38    & 24 & 93 & 49 & 32 & 100 & 67 & 54\\
   &m & 62 & 24    & 15 & 65 & 27 & 17 & 75 & 41 & 29\\
   &l & 61 & 23    & 14 & 62 & 25 & 15 & 72 & 37 & 25\\
   &e & 60 & 23    & 14 & 63 & 24 & 15 & 74 & 34 & 22\\
   \hline
   \multirow{4}{*}{PIONIER} & s & 75 & 33    & 21 & 85 & 41 & 27 & 97 & 63 & 51\\
  &m & 61& 24    & 15 & 64 & 26 & 16 & 71 & 39 & 27\\
   &l & 60 & 23 & 14 & 61 & 24 & 15 & 70 & 35 & 23\\
   &e & 59 & 23  & 14 & 61 & 23 & 14 & 71 & 32 & 20\\
\hline                  
\end{tabular}
\label{tab:single_inst}
\tablefoot{
Probabilities (in \%) are given for the small (s), medium (m), large (l), and extended (e) configurations 
\citep{VLTI_Manual_2024}, assuming a flux ratio $f = 1\%$ and projected distances within $r_{\text{max}}$.
}           
\end{table*}
\begin{table*}
\caption{
Maximum visibility deficit within the non-detection regions.
}
\centering
\begin{tabular}{c c|c c c|c c c|c c c}
\hline\hline

 &  & \multicolumn{3}{c|}{$T=0.1^\circ$} & \multicolumn{3}{c|}{$T=0.3^\circ$} & \multicolumn{3}{c}{$T=1^\circ$} \\
\cline{3-11}
\multicolumn{2}{c|}{Instrument \&}  & \multicolumn{3}{c|}{$r_{\text{max}}$ / au} &\multicolumn{3}{c|}{$r_{\text{max}}$ / au}&\multicolumn{3}{c}{$r_{\text{max}}$ / au} \\
\multicolumn{2}{c|}{configuration} & $0.1$ & $0.5$ & $1.0$ & $0.1$ & $0.5$ & $1.0$ & $0.1$ & $0.5$ & $1.0$ \\
\hline
   \multirow{4}{*}{MATISSE} & s & 0.6 & 1.7 & 2.0 & 0.8 & 1.8 & 2.0 & 0.8 & 2.0 & 2.0 \\
  & m & 0.9 & 1.3 & 1.3 & 1.2 & 2.0 & 2.0 & 2.0 & 2.0 & 2.0 \\
  & l & 0.6 & 0.8 & 0.9 & 1.2 & 1.8 & 1.8 & 2.0 & 2.0 & 2.0 \\
  & e & 0.6 & 0.6 & 0.6 & 1.9 & 1.9 & 1.9 & 2.0 & 2.0 & 2.0 \\
\hline
   \multirow{4}{*}{GRAVITY} & s & 0.6 & 2.0 & 2.0 & 1.2 & 2.0 & 2.0 & 1.6 & 2.0 & 2.0 \\
  & m & 0.9 & 0.9 & 0.9 & 1.8 & 2.0 & 2.0 & 2.0 & 2.0 & 2.0 \\
  & l & 0.4 & 0.9 & 0.9 & 1.2 & 1.6 & 1.6 & 1.9 & 2.0 & 2.0 \\
  & e & 0.6 & 0.6 & 0.6 & 1.9 & 1.9 & 1.9 & 1.9 & 2.0 & 2.0 \\
\hline
   \multirow{4}{*}{PIONIER} & s & 0.6 & 2.0 & 2.0 & 1.2 & 2.0 & 2.0 & 1.9 & 2.0 & 2.0 \\
  & m & 0.7 & 0.7 & 0.7 & 2.0 & 2.0 & 2.0 & 2.0 & 2.0 & 2.0 \\
  & l & 0.6 & 0.6 & 0.6 & 1.0 & 1.7 & 1.7 & 2.0 & 2.0 & 2.0 \\
  & e & 0.3 & 0.3 & 0.3 & 1.2 & 1.2 & 1.2 & 2.0 & 2.0 & 2.0 \\
\hline
\end{tabular}
\label{tab:single_dvmax}
\tablefoot{
The table lists the maximum visibility deficit, $\max|\Delta V|/f$, within regions where the closure phase satisfies $|\Phi| \le T$ and within a projected distance $r_{\text{max}}$. 
Values are given for three instruments and for the small (s), medium (m), large (l), and extended (e) configurations.
}
\end{table*}

Furthermore, we applied the same analysis to a combination of two simulated observations for the MATISSE medium configuration within a single night to assess the impact of rotating baselines during one night. As expected, the non-detection probabilities are lower in this case due to a more complete uv-coverage, although they remain non-zero and the blind-spots remain present.

Lastly, we combined different instruments using the medium and large configurations and calculated the non-detection probabilities for all possible combinations (see Table \ref{tab:all_comb}). Our results demonstrate that combining two or more of these instruments or configurations provides little advantage in reducing the non-detection probabilities of faint companions for detection accuracies up to $0.3^{\circ}$. Instead, the probabilities remain nearly equivalent to that of the best instrument-configuration-pair for all listed combinations. Larger differences in the non-detection probabilities were only found for a threshold of $1^{\circ}$. In this case, for companions within $0.1\,$au, the non-detection probability decreases to a minimum of 62\% by combining multiple configurations, while the best result of a single instrument with one configuration (i.e., PIONIER with the large configuration) amounts to 70\%. For the companions within $1\,$au, the values of the non-detection probability for all combined instruments with medium or large configurations were found to be roughly 15--23\,\%. We note, however, that companions located within 1\,au will exhibit rapid positional changes. Therefore, when combining two or more instruments, it would be essential to additionally account for their orbital motion and time-dependent variations.

\subsection{Case study of $\kappa$ Tuc A}
To illustrate the potential impact of a companion mimicking the signature of hot exozodiacal dust, we utilized observational data for $\kappa$~Tuc~A from a previously published study by \citet{Kirchschlager2020}. This system is of particular interest since it has been announced in \citet{2020AJ....159..265T} that there is an unknown companion derived from astrometric measurements \citep{2018ApJS..239...31B,2019ApJS..241...39B}. Our analysis is aimed at showcasing the possibility of a missed companion. However, we note that it does not serve as a complete analysis of the $\kappa$~Tuc~A observations.

For our case study, we therefore restricted the analysis to $\chi^2$ maps, which offer straightforward interpretability. While MCMC methods are well suited for parameter estimation, they often require advanced sampling strategies (e.g., tempered or parallel tempering) to avoid local minima. In contrast, grid searches combined with $\chi^2$ fitting are a well established approach in long–baseline interferometry. It was applied, for instance, in PIONIER surveys \citep[e.g.][]{2011A&A...535A..68A,Ertel2014} and in the CANDID algorithm \citep[e.g.][]{2015A&A...579A..68G}. A full MCMC treatment of all available $\kappa$~Tuc~A observations is left to the work of \citet{Stuber2025}.

\begin{figure*}
    \resizebox{\hsize}{!}{
        \includegraphics[clip]{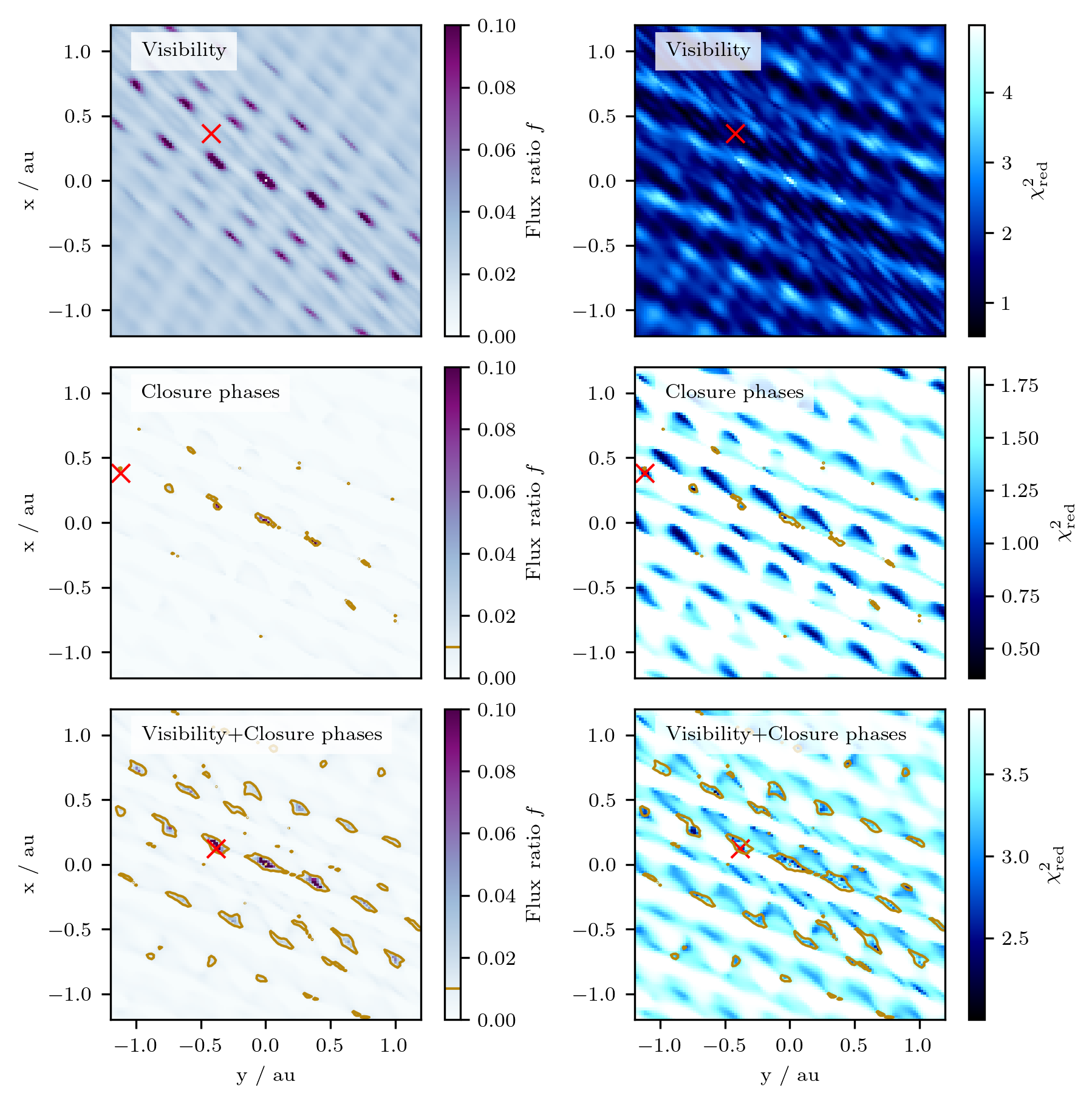}
    }
    \caption{Left: Best fit for companion-to-star flux ratio, $f,$ for measurements of $\kappa$~Tuc~A. Right: $\chi^2_{\rm red}$ for the position-dependent best-fit flux ratio. Each row corresponds to the different data used for fitting: visibility measurements (top), closure phases measurements (middle), and both visibility and closure phase measurements (bottom). Contour line in the middle and bottom plot mark the regions, where the flux ratio exceeds 1\%. Red crosses mark the pixel position corresponding to the minimum $\chi^2_{\rm red}$ value in each case.}
    \label{fig:kappa_tuc_fit_maps}
\end{figure*}

\begin{figure}
    \resizebox{\hsize}{!}{
        \includegraphics[clip]{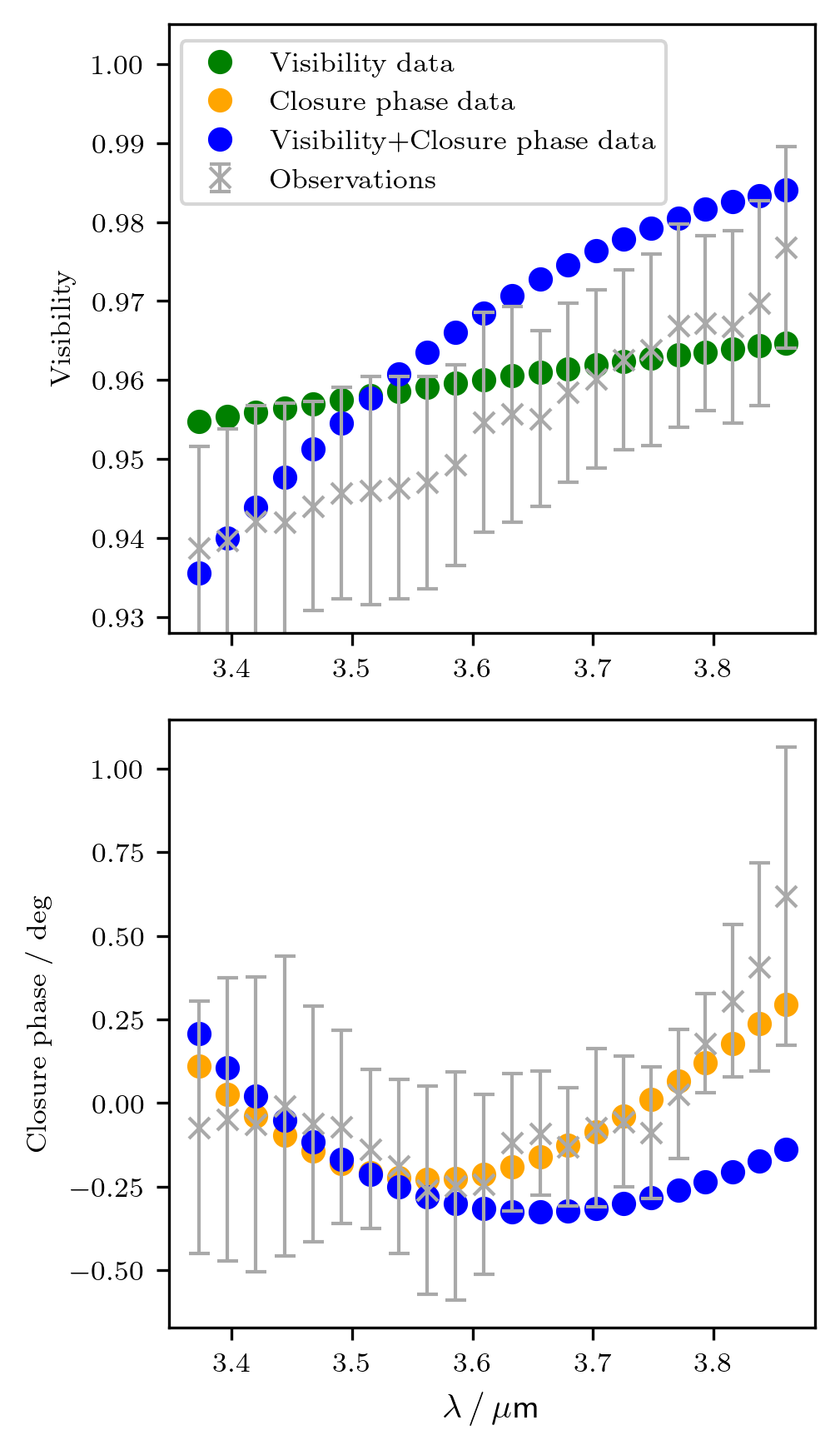}
    }
    \caption{Example best fits shown for one baseline (top) and one telescope triangle (bottom) as a function of the observing wavelength. The legend specifies the measured data used for the corresponding fit.}
    \label{fig:kappa_tuc_fits}
\end{figure}

The interferometric data were acquired with ATs on July 9 and July 11, 2019, using the MATISSE instrument in its medium configuration (K0-G2-D0-J3). We used observations for wavelengths between $3.37$ and $3.85\,\mu$m to cover the region that shows a significant visibility deficit. We modeled the system with a limb-darkened star in the center and a companion as a point source using~Eqs. \eqref{eq:v_ss_eq_A} and~\eqref{eq:closure_phases_general_eq}. We fit it separately to the measured visibilities and closure phases, followed by a simultaneous fit to both observables. Since a close-in companion could change its location significantly during the span of two days, we only considered the observations acquired on July 9, 2019. As a free parameter, we used the companion-to-star flux ratio $f\in\left[0,10\right]\,\%$ and calculated for different companion positions the corresponding value of the weighted reduced $\chi^2$, $\chi^2_{\rm red}$. Figure~\ref{fig:kappa_tuc_fit_maps} shows the results of our fitting procedures. Here, we minimized $\chi^2_{\rm red}$ pixelwise with regard to $f$, where each pixel represents the position of the companion in the field of view, up to a maximum projected separation of about 7\,au (i.e., 0.3\,as), which is larger than the region depicted in Fig.~\ref{fig:kappa_tuc_fit_maps}.
For each underlying dataset, the best-fit results are summarized in Table~\ref{tab:fit_results} and marked with the red crosses in Fig. \ref{fig:kappa_tuc_fit_maps}. 
\begin{table}
\caption{
Best-fit parameters for $\kappa$~Tuc~A derived from different data sets.
}
\centering
\begin{tabular}{c|c|c|c}
\hline\hline
Fitted data & $\chi^2_{\rm red}$ & $f$ & $r$\,/\,au\\
\hline
Visibility & 0.53& 0.03 & 0.55\\
Closure phases  & 0.36 & 0.007 & 1.2\\
Visibility+closure phases & 2.0 & 0.1 & 0.4\\
\hline
\end{tabular}
\label{tab:fit_results}
\tablefoot{
Fits were obtained using visibility, closure phase, and combined data sets. 
The corresponding projected separation is calculated.
}
\end{table}
We find that fitting the visibilities or closure phases individually yields very low $\chi^2_{\rm red}$ values (top right and middle right plots, respectively). However, the $\chi^2_{\rm red}$ value increases when combining both data sets (bottom right). Nonetheless, for the best fit, it still remains lower than the value reported by \cite{Kirchschlager2020}, where the measurements were explained by the presence of hot exozodiacal dust that has been modeled as a narrow ring. However, we note that the reported $\chi^2_{\rm red}$ cannot easily be compared, as they were computed based on models that differ in their number of free parameters and in the number of fitted observations. The plots furthermore reveal several spots with strong drops in $\chi^2_{\rm red}$. On the map of the best-fit flux ratio based on the observed closure phases (middle left), many regions correspond to cases with no companion (i.e., a flux ratio of 0), but the overall best-fit position is notably different, exhibiting a clearly non-zero flux. In combination with our determined very low value of $\chi^2_{\rm red}=0.36$, this finding suggests the presence of a companion. The higher $\chi^2_{\rm red}$ in the combined fit (bottom right), however, indicates that a single companion is unlikely to fully explain the entire system, suggesting the possibility of an additional source like hot exozodiacal dust or further companions. Additionally, the best fits for an arbitrarily selected baseline and one telescope triangle are shown exemplarily in Fig.~\ref{fig:kappa_tuc_fits}, showing a good agreement between the fitting results and the observations. Finally, using the closure phase fit results from Tables~\ref{tab:single_inst} and~\ref{tab:fit_results}, we estimate a non-detection probability. Assuming $f=0.007$ and a threshold value equal to the size of the error bar of the used closure phase measurements (i.e., $\approx 0.25^\circ$) gives a non-detection probability for the MATISSE medium configuration of $\sim21\,\%$ for the determined best-fit companion, which is large enough to potentially explain the reported variability and occasional non-detection of a visibility deficit of $\kappa$~Tuc~A.

To further illustrate the ambiguities in our analysis, we identified a secondary minimum in the $\chi^2_{\rm red}$-map of the closure phase fit, with a value of 0.38. This would correspond to a flux ratio of $f = 0.04$ at a projected separation of 1.6\,au. The non-detection probability for such a companion, given a detection threshold of approximately $0.25^\circ$, would be $\sim38\,\%$.

\section{Discussion}
\label{sec:discussion}
The hypothesis that a planetary companion might be responsible for the observed visibility deficit has  been dismissed in the context of different systems in the past \citep[e.g.,][]{2011A&A...534A...5D,Kirchschlager2020}. Such rejections were typically based on estimates of the Gaussian-distributed flux required to account for the deficit or on the argument that the observed closure phase signals were too weak to support a companion scenario. However, more recent studies have identified exoplanets or exoplanet candidates within projected separations of 1\,au in systems exhibiting hot exozodiacal dust signatures \citep{2017A&A...605A.103F, 2017AJ....154..135F,2019A&A...623A..72K}.
Based on these new discoveries, we reevaluated the companion hypothesis and argue that the assumptions underlying its early rejection may not have been universally valid. 
\paragraph*{Gaussian-distributed flux estimation:}
The first hot exozodiacal dust hypotheses were made on the basis of the observations without closure phase measurements as they used two telescope interferometers \citep[e.g.,][]{2006via..conf..251A,2009ApJ...704..150A}. These studies excluded a companion by calculating the Gaussian-distributed dust-to-star flux ratio that is necessary to produce the observed visibility deficit. 
The Gaussian visibility is a positive real number with zero imaginary part given by
\begin{equation}
    V_{\text{Gauss}}=\exp \left(-\frac{\pi^2\Theta^2_{\text{FOV}}B^2}{4\lambda^2\ln 2} \right)
    \label{Gauss}
,\end{equation}
with the field of view of the instrument, $\Theta_{\text{FOV}}$. This visibility value can take values from zero to one, which means that the visibility deficit of a system consisting of a star and Gaussian-spread dust is limited by $\left\lvert\Delta V_{\text{s+G}}(f)\right\lvert\leq f$ (analogously to Eq. \eqref{eq:dv_f}, we used the Gaussian visibility, $V_{\rm Gauss}$, instead of the real part of the visibility of a companion, $\mathfrak{Re}\left\{{\mathcal{V}_{\text{c}}}\right\}$). The derived upper limit of a companion in Eq.~\eqref{eq:max_delta_v} is, however, higher for the same assumed flux ratio between the additional flux source and the host star by a factor of two. This shows that the companion-induced visibility deficit is not necessarily well approximated by a Gaussian-distributed dust model within the field of view of a telescope. In other words, we find that the flux required to produce the measured visibility deficit is lower than previously assumed in studies excluding companions as a result of a flux estimate that is based on Gaussian-distributed dust \citep[e.g.,][]{2009ApJ...704..150A}. This discrepancy suggests that it may be worth reconsidering the previously dismissed companion hypothesis in case a Gauss estimation was applied. 
\paragraph*{Small closure phases:} 
In Sect.~\ref{sec:non_detection}, we calculated the probabilities for a non-detection of a faint companion based on closure phases and showed that, depending on its projected position, they can reach near-zero values and still affect the visibility deficit close to their theoretical upper limit derived in Eq.~\eqref{eq:max_delta_v}. Moreover, the hot exozodiacal dust phenomenon is typically associated with visibility deficits of at most a few percent \citep[except for 5 to 7\% for $\kappa$~Tuc~A in][]{Kirchschlager2020}, which is at an order that indicates it could be caused by faint companions. Therefore, a reliable rejection of the companion hypothesis requires an observational data set of sufficient uv-coverage that cannot be achieved by a single observation, enabling the exclusion of the companion residing inside blind spots during the observations (compare with Fig.~\ref{fig:dvs_with_mask}). We conclude that near-zero closure phases during a single observation do not necessarily exclude the presence of one or more companions, which has been done in past studies \citep[e.g.,][]{Kirchschlager2020}. 
Moreover, based on the Tables~\ref{tab:single_inst} and~\ref{tab:single_dvmax}, the observations made with the small configuration, or short baselines, could easily lead to false non-detections, given its high non-detection probability in closure phases combined with the highest possible visibility deficit.
Performing a reanalysis of observations where the companion hypothesis was rejected based on small measured closure phases is crucial and may reshape our understanding of the observed visibility deficit. 
\paragraph*{Distinguishing faint companions from hot exozodiacal dust:} 
Altogether, these findings further highlight the importance of robust diagnostic methods to detect hot dust and distinguish it from faint companions. Multiple observations at different wavelengths are necessary to decrease the non-detection probability, as different wavelengths correspond to different blind spot patterns in the closure phases. 

Based on the insights gained from Tables~\ref{tab:single_inst} and~\ref{tab:single_dvmax}, we propose a novel, additional criterion that can help us to distinguish between these two phenomena in future analyses of observed visibility deficits. This method assumes near-zero closure phases for hot exozodiacal dust and requires an estimate for the maximum possible flux of a potential companion (see Sect.~\ref{sec:theoretical_limits}). The following description is valid for the example $\kappa$~Tuc~A observations, but can be applied to any other system by calculating the corresponding probabilities as described before. 

If non-zero closure phases are detected above the assumed threshold, the presence of a companion is the most plausible explanation. If the closure phases are below the threshold and the maximum observed visibility deficit exceeds the maximum value from Table~\ref{tab:single_dvmax}, the companion as the primary source can be excluded with a high probability, indicating that the observed signal likely originates from hot exozodiacal dust. Conversely, if the maximum visibility deficit is lower than the highest value in Table~\ref{tab:single_dvmax} for the selected threshold, the result is inconclusive and  additional observation data are required. This means that the two phenomena cannot be distinguished based on the current measurements. Furthermore, it is important to note that the observed signal could also arise from other asymmetric structures, such as dust clumps or inhomogeneities in a dust ring. These possibilities should be considered separately when interpreting the data to avoid misattributing the signal to a companion or the so far often assumed azimuthally homogeneous ring-shaped distribution of the hot exozodiacal dust.
\paragraph*{Caveats:} 
To analyze the differences between configurations, we used fixed baseline lengths and constant orientations relative to the position of the assumed companion. In practice, their values will vary for different observations depending on the current position of the observed object, especially if it has a short orbital period of weeks or even days, and on the time of observation, which influences the resulting non-detection probabilities. As demonstrated in this study, baseline length is a critical factor in reducing non-detection probabilities in closure phases, as more extended configurations are associated with higher resolution and thus with lower probabilities (see Tables \ref{tab:single_inst} and \ref{tab:all_comb}). Furthermore, these calculated probabilities only take into account the different positions, but not the time that a companion on a non-circular orbit spends at each position; on an elliptical orbit a companion will spend more time at larger distances compared to smaller distances. The results further depend on the exact properties of the orbit of the companion such as the inclination, the position angle, the radial distance between the companion and the star, the distance to the host star, which influences its angular size and visibility (see Eq.~\ref{eq:star_vis}), and the companion-to-star flux ratio. In this study, we used the same companion-to-star flux ratio for each wavelength, as the bandwidth is relatively narrow and no significant variations are expected within this range. Altogether, making a case-specific statement regarding the non-detection probabilities of a system therefore requires an in-depth analysis. To assist researchers in planning their observations, we have made an accompanying Jupyter notebook publicly available alongside this paper (see details in Sect.~\ref{sec:summary}). 

\section{Summary and conclusions}
\label{sec:summary}
In this study, we investigated the hot exozodiacal dust hypothesis, proposing that some of the phenomena previously attributed to hot dust might, in fact, be caused by faint companions, despite their exclusion in past analyses. We derived theoretical upper limits in Sect.~\ref{sec:theoretical_limits} for companion-induced detected visibility deficits and closure phases, assuming a limb-darkened host star that is observed with the instrument PIONIER, GRAVITY, or MATISSE at the VLTI. 
Our analysis indicates that these potential companions cannot be easily dismissed and could offer a simpler, more comprehensive explanation for certain observed features that are commonly attributed to the presence of hot exozodiacal dust. 
Additionally, we reevaluated one previous observation of $\kappa$~Tuc~A. We report the possibility of a faint companion with a companion-to-star flux ratio of 0.7\,\% and a non-detection probability of  $\sim21\,\%$. However, to draw clear conclusions, a more comprehensive analysis incorporating all available observations of $\kappa$~Tuc~A is required. However, this does not preclude the relevance of the hot dust hypothesis, as both mechanisms could contribute to the observed phenomena, which include:
\begin{itemize}
    \item the observed visibility deficits, which can be influenced by companions in a way that leads to larger deficits, even with lower companion fluxes compared to Gaussian-distributed hot exozodiacal dust (Sect.~\ref{sec:discussion}).
    \item low closure phases, which  our analysis has shown can occur and could lead to significant non-detection probabilities for companions due to configuration-specific blind spots (Sect.~\ref{sec:results}). This implies that detecting small closure phases in a single observation is neither a sufficient criterion for a rejection nor a necessary one (Sect.~\ref{sec:discussion}), even in the case of a significant detected visibility deficit.
    \item The observed variability of the exozodiacal dust, for which a definitive explanation is still lacking, but which could be accounted for by the orbital motion of companions. 
    \item Significant levels of infrared flux excesses associated with the presence of small, hot dust grains near the host star, which might be better understood originating from a companion. 
\end{itemize}
Altogether, our findings emphasize the importance of rigorously reevaluating past detections with careful consideration of both hot dust and faint companions. For that purpose, we have proposed a novel method of distinguishing between both origins (presented in Sect.~\ref{sec:discussion}), which is based on visibility deficits and closure phases estimations.
Accurately characterizing hot exozodiacal dust is especially crucial for understanding the innermost regions of planetary systems, refining detection strategies, and retrieving signals from close-in exoplanets. In this context, understanding non-detection probabilities becomes essential, as case-specific analyses are key to interpreting the presence or absence of companions. To assist with this pursuit, we have made a Jupyter notebook publicly available alongside this paper, helping researchers evaluate existing observations and plan future ones.

\begin{acknowledgements}
      This work was supported by a federal state scholarship of Kiel University. We thank the organizers and participants of the 12th VLTI School of Interferometry. This work has made use of data from the European Space Agency (ESA) mission
{\it Gaia} (\url{https://www.cosmos.esa.int/gaia}), processed by the {\it Gaia}
Data Processing and Analysis Consortium (DPAC,
\url{https://www.cosmos.esa.int/web/gaia/dpac/consortium}). Funding for the DPAC
has been provided by national institutions, in particular the institutions
participating in the {\it Gaia} Multilateral Agreement. This research has made use of the Jean-Marie Mariotti Center \texttt{Aspro2} service \footnote{Available at http://www.jmmc.fr/aspro}. FK has received funding from the European Research Council (ERC) under the European Union's Horizon 2020 research and innovation programme DustOrigin ERC-2019-StG-851622. TAS has received financial support from the National Aeronautics and Space Administration under grants 80NSSC23K0288 (PI: Faramaz) and 80NSSC23K1473 (PI: Ertel).
\end{acknowledgements}

\bibliography{literature}
\bibliographystyle{aa}

\begin{appendix}
\section{Derivation of faint companion signature}
\label{app:deriv}
\begin{figure}[h]
    \resizebox{\hsize}{!}{
        \includegraphics[clip]{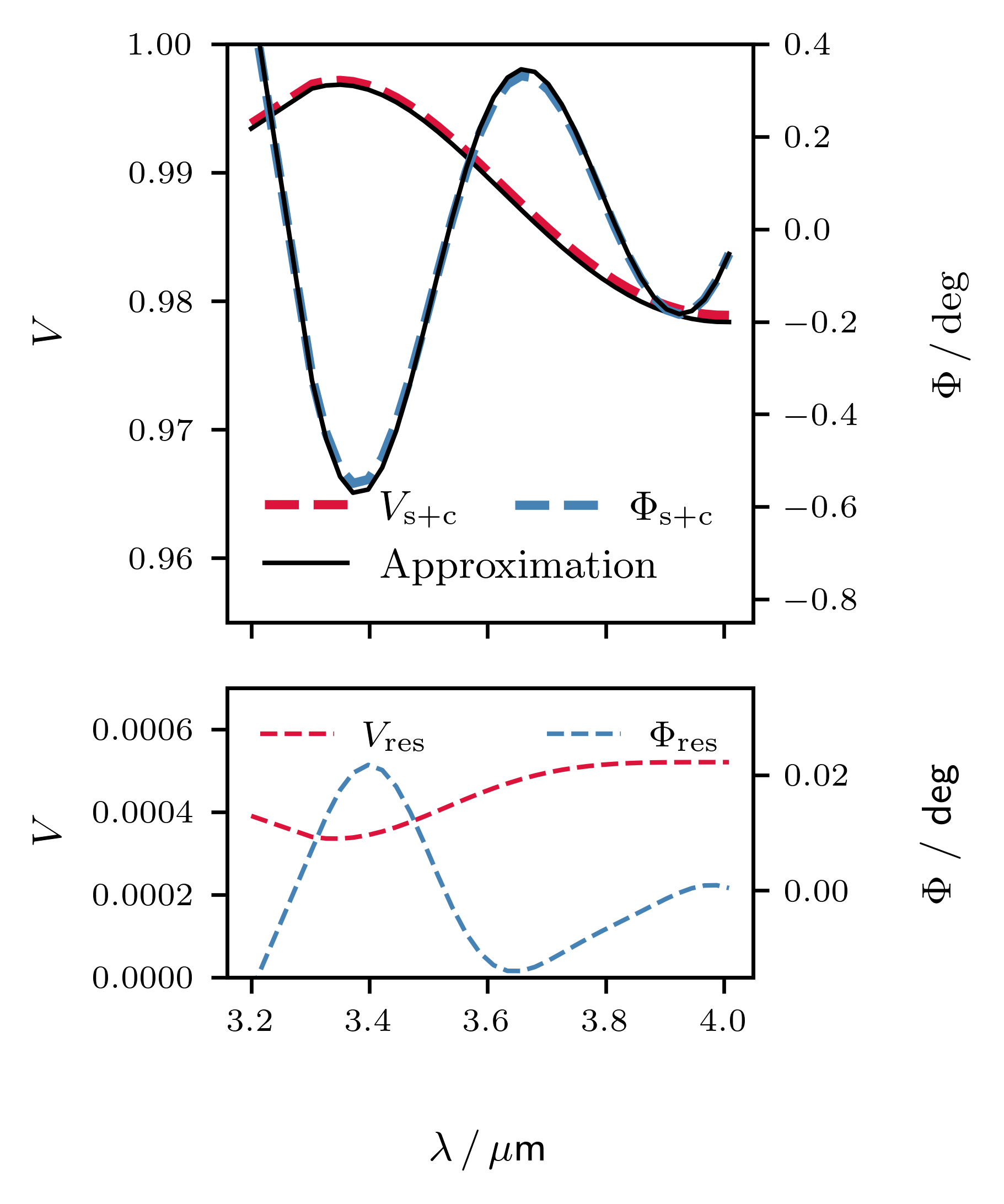}
    }
    \caption{Numerical test for the linear approximations (solid black lines), given in Eqs. \eqref{eq:taylor_sys} and \eqref{eq:CP}. Top: Expected visibility amplitudes $V_{\texttt{s+c}}$ and closure phases $\Phi_{\texttt{s+c}}$ of a binary system composed of a central star and a companion (dashed lines), exemplarily modeled for $\kappa$~Tuc~A  and a faint companion at the projected distance of $\approx 1.4\,$au, assuming a companion-to-star flux ratio of $f=1\%$ (for details refer to Fig.~\ref{fig:cps_and_vis_example}). Bottom: Residuals (res) between the exact model and the approximation, shown for the visibility and closure phase. }
    \label{fig:num_test}
\end{figure}

Given the total visibility of the system
\begin{equation*}
    \mathcal{V}_{\text{s+c}}= \frac{V_{\text{s}}+\mathcal{V}_{\text{c}}f}{1+f},
\end{equation*}
its real part
\begin{equation*}
    \mathfrak{Re}\left\{\mathcal{V}_{\text{s+c}}\right\}= \frac{V_{\text{s}}+\mathfrak{Re}\left\{\mathcal{V}_{\text{c}}\right\}f}{1+f}
\end{equation*}
and its imaginary part
\begin{equation*}
    \mathfrak{Im}\left\{\mathcal{V}_{\text{s+c}}\right\}= \frac{\mathfrak{Im}\left\{\mathcal{V}_{\text{c}}\right\}f}{1+f}
\end{equation*}
can be used to derive the total visibility amplitude $V_{\text{s+c}}$ as 
\begin{eqnarray}
    &V_{\text{s+c}} = \sqrt{\mathfrak{Re}\left\{\mathcal{V}_{\text{s+c}}\right\}^2 + \mathfrak{Im}\left\{\mathcal{V}_{\text{s+c}}\right\}^2}\\
    =& \frac{1}{1+f}\sqrt{V_s^2 + 2V_s\mathfrak{Re}\left\{\mathcal{V}_{\text{c}}\right\}f + \left(\mathfrak{Re}\left\{\mathcal{V}_{\text{c}}\right\}^2 + \mathfrak{Im}\left\{\mathcal{V}_{\text{c}}\right\}^2 \right)f^2}\\
    =& \frac{1}{1+f}\sqrt{V_s^2 + 2V_s\mathfrak{Re}\left\{\mathcal{V}_{\text{c}}\right\}f + f^2}.
\end{eqnarray}
To derive the last equation we made use of the relation $\mathfrak{Re}\left\{\mathcal{V}_{\text{c}}\right\}^2 + \mathfrak{Im}\left\{\mathcal{V}_{\text{c}}\right\}^2 = 1$. 
We find that the entire expression is independent on the imaginary part of the complex visibility of the companion. \\
As the companion is faint, we can make the Taylor series at $f=0$:
    \begin{align}
V_{\text{s+c}} &= V_{\text{s+c}}\,(f = 0) + V'_{\text{s+c}}\,(f = 0)f + \mathcal{O}(f^2) \\
              &= V_{\text{s}} + (\mathfrak{Re}\left\{\mathcal{V}_{\text{c}}\right\} - V_{\text{s}})f + \mathcal{O}(f^2).
    \end{align}
The comparison between the exact analytical calculation and the derived linear approximation is shown for an example in Fig.~\ref{fig:num_test}.

\onecolumn
\section{Visibility and closure phase maps}
\label{sec:pionier_gravity}
\begin{figure*}[!htb]
    \resizebox{\hsize}{!}{
        \includegraphics[clip]{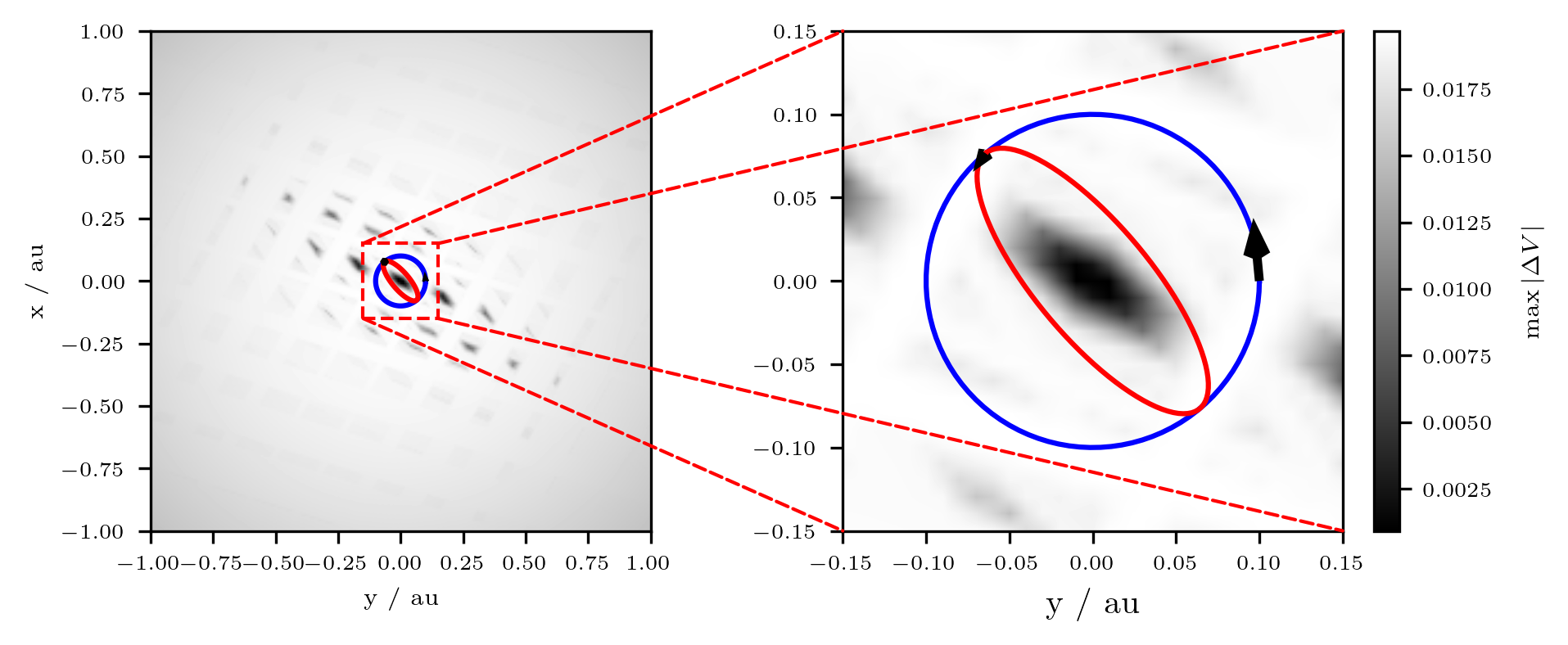}
    }
    \caption{Maximum visibility deficit of the system in dependence on the position of the assumed companion for all simulated PIONIER wavelengths and medium configuration. Field of view of \( 2 \, \mathrm{au} \times 2 \, \mathrm{au}  \) (top) and \( 0.3 \, \mathrm{au} \times 0.3 \, \mathrm{au}  \) (bottom). The arrows show the direction of the orbiting companion.}
    \label{fig:max_dvs_pionier}
\end{figure*}
\begin{figure*}[!htb]
    \resizebox{\hsize}{!}{
        \includegraphics[clip]{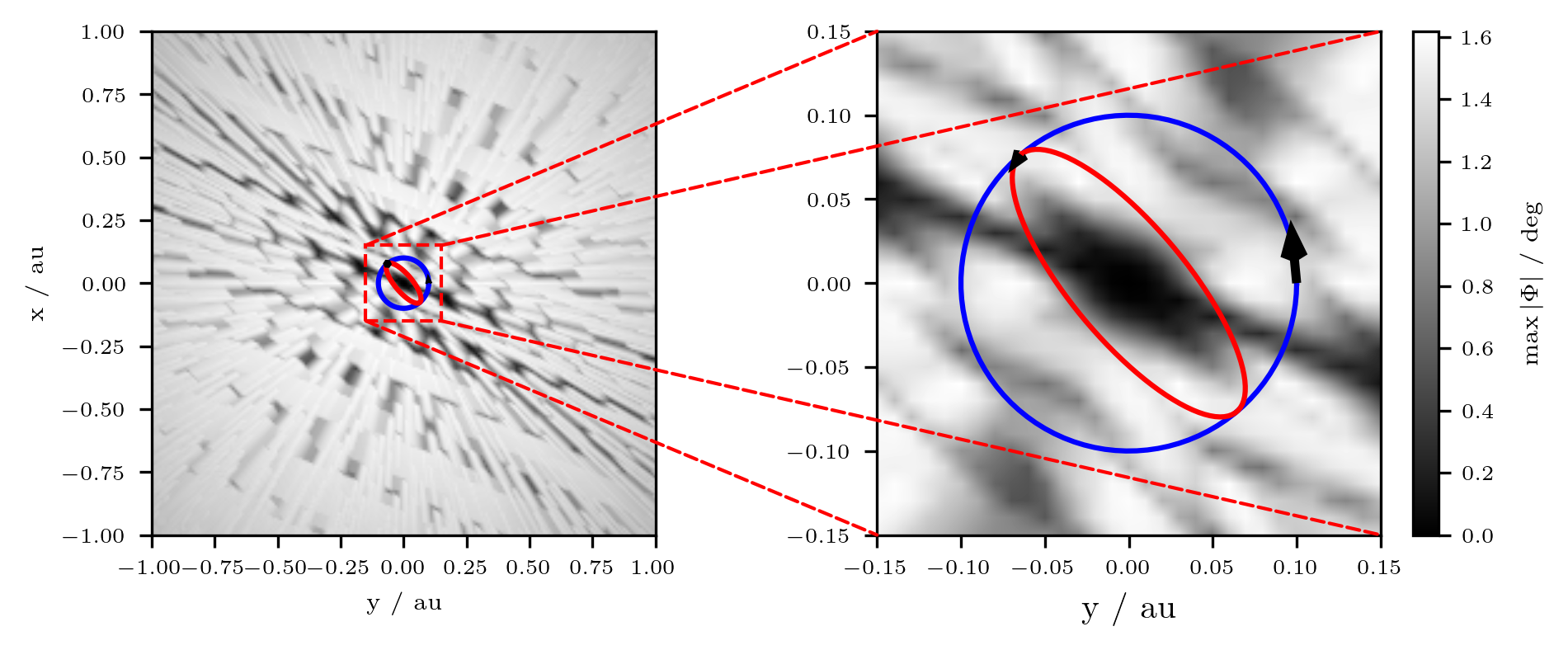}
    }
    \caption{Maximum closure phase of the system in dependence on the position of the assumed companion for all simulated PIONIER wavelengths and medium configuration. Field of view of \( 2 \, \mathrm{au} \times 2 \, \mathrm{au}  \) (top) and \( 0.3 \, \mathrm{au} \times 0.3 \, \mathrm{au}  \) (bottom). The arrows show the direction of the orbiting companion.}
    \label{fig:max_cps_pionier}
\end{figure*}
\begin{figure*}
    \resizebox{\hsize}{!}{
        \includegraphics[clip]{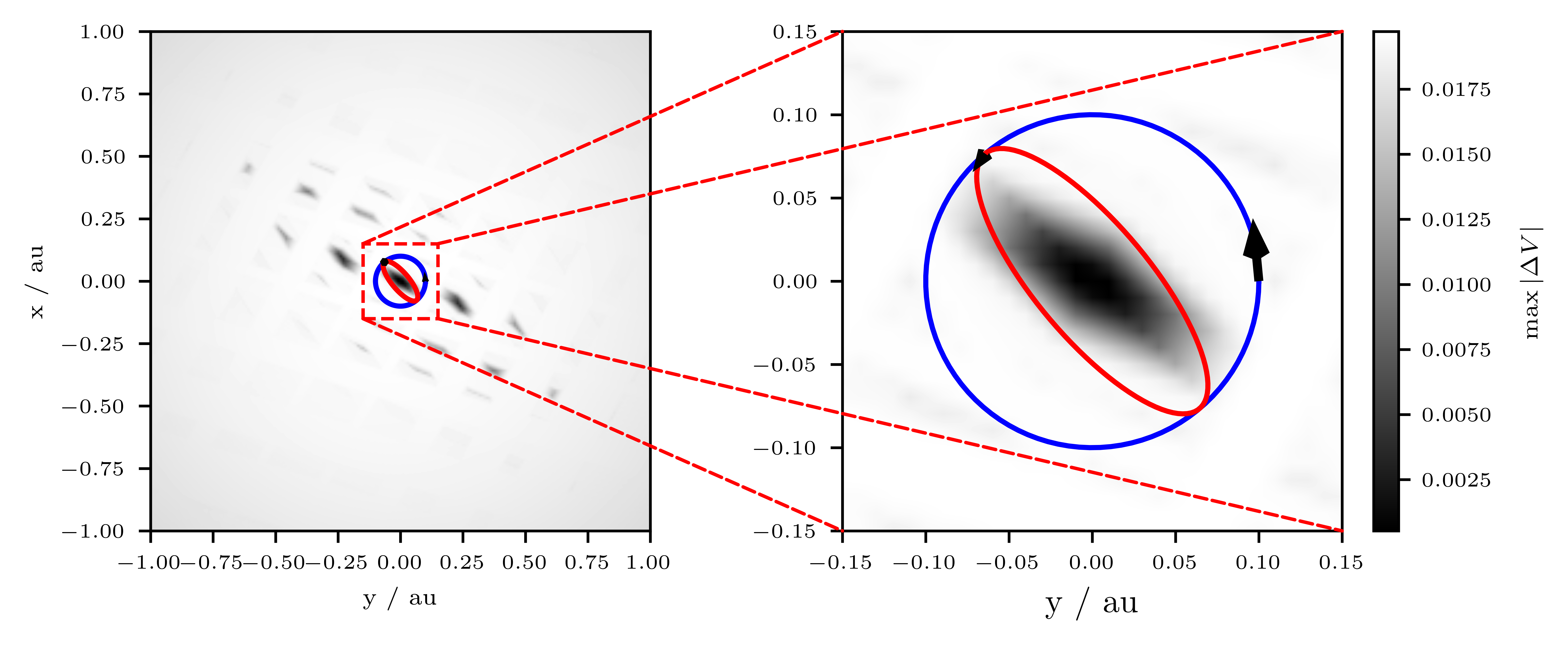}
    }
    \caption{Maximum visibility deficit of the system in dependence on the position of the assumed companion for all simulated GRAVITY wavelengths and medium configuration. Field of view of \( 2 \, \mathrm{au} \times 2 \, \mathrm{au}  \) (left) and \( 0.3 \, \mathrm{au} \times 0.3 \, \mathrm{au}  \) (right). The arrows show the direction of the orbiting companion.}
    \label{fig:max_dvs_gravity}
\end{figure*}
\begin{figure*}
    \resizebox{\hsize}{!}{
        \includegraphics[clip]{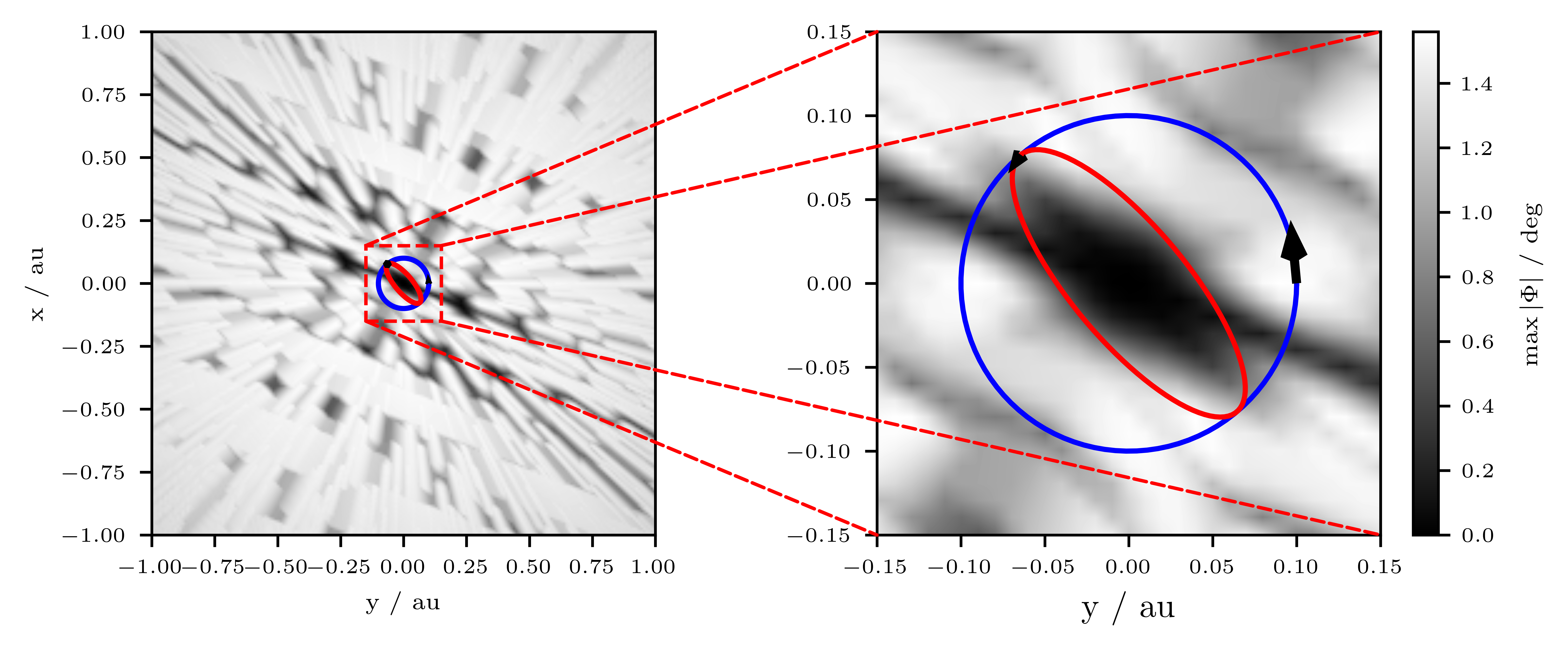}
    }
    \caption{Maximum closure phase of the system in dependence on the position of the assumed companion for all simulated GRAVITY wavelengths and medium configuration. Field of view of \( 2 \, \mathrm{au} \times 2 \, \mathrm{au}  \) (left) and \( 0.3 \, \mathrm{au} \times 0.3 \, \mathrm{au}  \) (right). The arrows show the direction of the orbiting companion.}
    \label{fig:max_cps_gravity}
\end{figure*}
\onecolumn
\section{Non-detection probabilities}
\begin{table*}[h!]
\caption{
Non-detection probabilities for instrument combinations and configurations.
}
\centering
\small
\begin{tabular}{l| l l l| l l l| l l l}
\hline
Combination & \multicolumn{3}{|c|}{Threshold = 0.1$^\circ$} & \multicolumn{3}{|c|}{Threshold = 0.3$^\circ$} & \multicolumn{3}{|c}{Threshold = 1$^\circ$} \\
\cline{2-10}

 & 0.1 au & 0.5 au & 1 au & 0.1 au & 0.5 au & 1 au & 0.1 au & 0.5 au & 1 au \\ \hline
Gl Mm & 61 & 23 & 14 & 62 & 24 & 15 & 72 & 31 & 19 \\
Gl Ml & 61 & 23 & 14 & 62 & 24 & 15 & 68 & 29 & 19 \\
Gm Gl & 61 & 23 & 14 & 62 & 24 & 15 & 68 & 28 & 18 \\
Gm Ml & 61 & 24 & 15 & 63 & 25 & 15 & 69 & 30 & 19 \\
Gm Mm & 62 & 24 & 15 & 65 & 26 & 16 & 74 & 35 & 23 \\
Mm Ml & 62 & 24 & 15 & 65 & 26 & 16 & 75 & 34 & 22 \\
Pl Ml & 60 & 23 & 14 & 61 & 23 & 14 & 65 & 27 & 17 \\
Pl Gm & 60 & 23 & 14 & 61 & 23 & 14 & 66 & 27 & 17 \\
Pl Mm & 60 & 23 & 14 & 61 & 23 & 14 & 67 & 28 & 18 \\
Pl Gl & 60 & 23 & 14 & 61 & 23 & 14 & 65 & 27 & 17 \\
Pm Gl & 60 & 23 & 14 & 61 & 23 & 14 & 65 & 27 & 17 \\
Pm Mm & 61 & 23 & 14 & 63 & 25 & 16 & 68 & 31 & 20 \\
Pm Ml & 61 & 23 & 14 & 62 & 24 & 15 & 67 & 29 & 19 \\
Pm Gm & 61 & 23 & 14 & 63 & 25 & 16 & 68 & 31 & 20 \\
Pm Pl & 60 & 23 & 14 & 61 & 23 & 14 & 64 & 26 & 17 \\
Gl Mm Ml & 61 & 23 & 14 & 62 & 24 & 15 & 67 & 27 & 17 \\
Gm Gl Ml & 61 & 23 & 14 & 62 & 24 & 15 & 65 & 26 & 16 \\
Gm Mm Ml & 61 & 24 & 15 & 63 & 25 & 15 & 69 & 29 & 18 \\
Gm Gl Mm & 61 & 23 & 14 & 62 & 24 & 15 & 68 & 28 & 17 \\
Pl Gm Mm & 60 & 23 & 14 & 61 & 23 & 14 & 66 & 26 & 16 \\
Pl Gl Mm & 60 & 23 & 14 & 61 & 23 & 14 & 65 & 26 & 16 \\
Pl Gm Ml & 60 & 23 & 14 & 61 & 23 & 14 & 64 & 25 & 16 \\
Pl Gm Gl & 60 & 23 & 14 & 61 & 23 & 14 & 64 & 25 & 16 \\
Pl Gl Ml & 60 & 23 & 14 & 61 & 23 & 14 & 64 & 26 & 16 \\
Pl Mm Ml & 60 & 23 & 14 & 61 & 23 & 14 & 65 & 26 & 16 \\
Pm Pl Gm & 60 & 23 & 14 & 61 & 23 & 14 & 64 & 26 & 16 \\
Pm Gl Mm & 60 & 23 & 14 & 61 & 23 & 14 & 65 & 26 & 16 \\
Pm Gm Gl & 60 & 23 & 14 & 61 & 23 & 14 & 65 & 26 & 16 \\
Pm Gm Ml & 61 & 23 & 14 & 62 & 24 & 15 & 66 & 27 & 17 \\
Pm Mm Ml & 61 & 23 & 14 & 62 & 24 & 15 & 66 & 27 & 17 \\
Pm Gl Ml & 60 & 23 & 14 & 61 & 23 & 14 & 64 & 25 & 16 \\
Pm Gm Mm & 61 & 23 & 14 & 63 & 25 & 15 & 68 & 30 & 19 \\
Pm Pl Mm & 60 & 23 & 14 & 61 & 23 & 14 & 64 & 25 & 16 \\
Pm Pl Gl & 60 & 23 & 14 & 61 & 23 & 14 & 63 & 25 & 15 \\
Pm Pl Ml & 60 & 23 & 14 & 61 & 23 & 14 & 63 & 25 & 15 \\
Gm Gl Mm Ml & 61 & 23 & 14 & 62 & 24 & 15 & 65 & 26 & 16 \\
Pl Gm Mm Ml & 60 & 23 & 14 & 61 & 23 & 14 & 64 & 25 & 16 \\
Pl Gm Gl Mm & 60 & 23 & 14 & 61 & 23 & 14 & 64 & 25 & 16 \\
Pl Gm Gl Ml & 60 & 23 & 14 & 61 & 23 & 14 & 63 & 25 & 15 \\
Pl Gl Mm Ml & 60 & 23 & 14 & 61 & 23 & 14 & 64 & 25 & 16 \\
Pm Pl Gm Ml & 60 & 23 & 14 & 61 & 23 & 14 & 63 & 25 & 15 \\
Pm Gm Mm Ml & 61 & 23 & 14 & 62 & 24 & 15 & 66 & 27 & 17 \\
Pm Pl Gl Mm & 60 & 23 & 14 & 61 & 23 & 14 & 63 & 24 & 15 \\
Pm Gm Gl Ml & 60 & 23 & 14 & 61 & 23 & 14 & 64 & 25 & 15 \\
Pm Pl Gm Mm & 60 & 23 & 14 & 61 & 23 & 14 & 64 & 25 & 16 \\
Pm Gl Mm Ml & 60 & 23 & 14 & 61 & 23 & 14 & 64 & 25 & 15 \\
Pm Pl Gl Ml & 60 & 23 & 14 & 61 & 23 & 14 & 62 & 24 & 15 \\
Pm Gm Gl Mm & 60 & 23 & 14 & 61 & 23 & 14 & 65 & 26 & 16 \\
Pm Pl Gm Gl & 60 & 23 & 14 & 61 & 23 & 14 & 63 & 24 & 15 \\
Pm Pl Mm Ml & 60 & 23 & 14 & 61 & 23 & 14 & 63 & 25 & 15 \\
Pl Gm Gl Mm Ml & 60 & 23 & 14 & 61 & 23 & 14 & 63 & 25 & 15 \\
Pm Pl Gl Mm Ml & 60 & 23 & 14 & 61 & 23 & 14 & 62 & 24 & 15 \\
Pm Pl Gm Mm Ml & 60 & 23 & 14 & 61 & 23 & 14 & 63 & 25 & 15 \\
Pm Gm Gl Mm Ml & 60 & 23 & 14 & 61 & 23 & 14 & 64 & 25 & 15 \\
Pm Pl Gm Gl Mm & 60 & 23 & 14 & 61 & 23 & 14 & 63 & 24 & 15 \\
Pm Pl Gm Gl Ml & 60 & 23 & 14 & 61 & 23 & 14 & 62 & 24 & 15 \\
Pm Pl Gm Gl Mm Ml & 60 & 23 & 14 & 61 & 23 & 14 & 62 & 24 & 15 \\
\hline
\end{tabular}
\label{tab:all_comb}
\tablefoot{
Probabilities (\%) are given for combinations of PIONIER (P), GRAVITY (G), and MATISSE (M) in medium (m) and large (l) configurations, assuming $f = 1\%$, for different thresholds and within specified radii.
}
\end{table*}
\end{appendix}

\end{document}